\documentclass[prd,preprint,superscriptaddress,amsmath,amssymb,nofootinbib]{revtex4}
\pdfoutput=1

\usepackage{graphicx}
\usepackage{dcolumn}
\usepackage{bm}
\usepackage{amssymb}
\usepackage{amsmath}
\usepackage{epsfig}    
\usepackage{color}
\usepackage{slashed}
\usepackage{hhline}
\usepackage{youngtab}
\usepackage{multirow}
\newcommand{\met}{\not{\!\!{ E}}_{T}}
\newcommand{\mpt}{\not{\! { p}}_{T}}

\def\be{\begin{equation}}
\def\ee{\end{equation}}
\newcommand{\bea}{\begin{eqnarray}}
\newcommand{\eea}{\end{eqnarray}}
\newcommand{\nn}{\nonumber}

\newcommand{\RNum}[1]{\uppercase\expandafter{\romannumeral #1\relax}}

\begin{document}

 \begin{flushright} {KIAS-P18110},  APCTP Pre2018-016  \end{flushright}

\title{A neutrino mass model with hidden $U(1)$ gauge symmetry}

\author{Haiying Cai}
\email{haiying.cai@apctp.org}
\affiliation{Asia Pacific Center for Theoretical Physics, Pohang, Gyeongbuk 790-784, Republic of Korea}

\author{Takaaki Nomura}
\email{nomura@kias.re.kr}
\affiliation{School of Physics, KIAS, Seoul 02455, Republic of Korea}

\author{Hiroshi Okada}
\email{hiroshi.okada@apctp.org}
\affiliation{Asia Pacific Center for Theoretical Physics, Pohang, Gyeongbuk 790-784, Republic of Korea}

\date{\today}

\begin{abstract} 
 We propose a realisation of inverse seesaw model controlled by hidden $U(1)$ gauge symmetry,  and discuss  the impact of a  bosonic dark matter (DM) candidate  by imposing a $Z_2$ parity.   We present the detail of  scalar spectra and  apply the Casas-Ibarra parametrisation to fit the neutrino oscillation data.  For this  model,   the allowed region is extracted to explain  the observed relic density and the muon $g-2$ discrepancy,   satisfying  flavor constraints  with DM involved.  We interpret the  DM annihilation into  $\bar f f $ including all SM charged fermions  and   investigate the direct detection  to place the  bound on  DM-Higgs coupling.  Finally the LHC  DM production  is explored in light  of  charged lepton pair signature plus  missing  transverse energy.  

 \end{abstract} 
\maketitle

\section{Introduction}
The discovery of  the Higgs boson  at the Large Hadron Collider (LHC) in 2012 announced the success of the Standard Model (SM) and  data collected so far have affirmatively validated the high precision of this framework in explaining most of phenomenology. However the deviations from the  expected SM prediction within  the current  experimental uncertainty  can still accommodate the possibility of new particles existence motivated by those interesting scenarios of  extra dimension, supersymmetry and composite Higgs model, etc.  There are also several aspects such as non-zero masses of neutrinos and dark matter (DM) candidate which should involve physics beyond the SM.
For neutrinos, several recent experiments observing the neutrino oscillations  confirmed that the neutrino has a  tiny mass  at the order $<  0.1$ eV,  which is much smaller compared with the SM quarks and leptons.   One favourable  neutrino model is  supposed to account for  the important features related to three active neutrinos with mixing angles  of  $\theta_{12, 13, 23}$ and the two neutrino mass square differences, $\Delta m_{12}^2$ and $|\Delta m_{23}^2|$  consistent with the observations ~\cite{Forero:2014bxa, Capozzi:2018ubv, Esteban:2018azc}. Furthermore,  several issues  are very poorly understood, including whether neutrino is a Dirac fermion or  Majorana one, in  normal  or inverted hierarchy  pattern for  the mass ordering, and the exact value of  CP violation phase, and so on.  In particular, the presence of Majorana  field violating the lepton number in this type of models  leads to the neutrinoless double beta decay detectable in experiments,  as well as  possibility to explain the Baryon Asymmetry of the Universe  via "leptogenesis"~\cite{Fukugita:1986hr}.  Thus it is  important to explore and analyze viable neutrino mass models in order to reveal  the nature and role of the neutrino sector. 

The simplest idea to realize a tiny neutrino mass is  seesaw mechanism by introducing  heavier neutral fermions which can obtain Majorana mass at the GUT scale $\sim 10^{15}$ GeV. There are several types of seesaw models  after  a long time of evolving, such as  type-I  seesaw (aka canonical seesaw) or type-III seesaw involving either a $SU(2)_L$  singlet or triplet right-handed neutral fermions ~\cite{Seesaw1, Seesaw2, Seesaw3, Seesaw4} .  One alternative mechanism to obtain a small mass is  the radiative seesaw provided the neutrino mass can only be generated at the one-loop level, such as the model proposed in the paper of \cite{Ma:2006km}, where the neutral $Z_2$ odd scalar  interacting with neutrino could be the DM candidate.  While an inverse seesaw  is  a promising scenario to reproduce neutrino masses and their mixings by introducing both left and right-handed neutral fermions so that the seesaw mechanism is proceeded via a two-step mediation.  This type of mechanism is often considered in extended  gauge models such as the superstring inspired model or  left-right models in unified gauge group~\cite{Mohapatra:1986bd, Wyler:1982dd}. In this paper, we propose an inverse seesaw model with several extra scalars charged by  a hidden U(1) symmetry,  which provides rather natural  hierarchies among active neutrinos and heavy neutral fermions even at the tree level compared to another similar scenario of  linear seesaw~\cite{Akhmedov:1995ip, Akhmedov:1995vm}. 
In analogy to the radiative seesaw,   an inert scalar is identified as  DM candidate in this model, whose  interaction with charged SM leptons and quarks will not only produce the observed relic density, but also give rise to rich LHC phenomenology.  The typical LHC  signature related to DM production is jets or leptons plus large missing energy, which  provides complimentary limits for the parameter space.  In fact  our scenario  allows certain advantage for the DM production  at the collider, since in order to obtain the LHC bound,  it is crucial to tag the accompanied SM particles like charged leptons in our case.

This letter is organised as follows. In Section~II, we present our model  by showing new particle fields and  symmetries, where  the inverse seesaw mechanism is implemented in a framework of hidden $U(1)$ gauge symmetry. We  add an inert boson that is expected to be a dark matter (DM) candidate, where a $Z_2$ symmetry is imposed to assure the stability of DM.  The scalar potential is constructed  to trigger spontaneous symmetry breaking and generate the required mass hierarchy.  In Section~III, We review the electroweak bounds from the lepton flavor violation processes  related to charged lepton and $Z$ boson decays. In particular, we provide an  analytic formula for the  annihilation  amplitude squared in terms of four momentum of DM and SM fermions,   verified  with the  chiral limit  result as  an expansion  of $v_{\rm rel}$ in the literature.  We  further interpret the impact from the observed relic density, along with the direct direction bound on a Higgs-portal term.  A numerical analysis is carried out  to search for the allowed parameter region.  In Section~IV,  we  discuss the LHC collider physics in our model by exploring  the pair production of vector-like charged leptons, which subsequently decay into the DM plus  SM leptons.  Finally we devote the Section~V to the summary and conclusion of our results.

\section{The Model}
\begin{table}[t!]
\begin{tabular}{|c||c|c|c|c||c|c|c|c|}\hline\hline  
& ~$U_R(U_L)$~&~ $D_R(D_L)$~  &~ $E_R(E_L)$~ & ~$N_R(N_L)$~ & ~$H'$~ & ~$\varphi$~ & ~$\varphi'$~& ~$\chi$~ \\\hline\hline 
$SU(3)_C$ & $\bm{3}$  & $\bm{3}$ & $\bm{1}$ & $\bm{1}$ & $\bm{1}$ & $\bm{1}$ & $\bm{1}$ & $\bm{1}$ \\\hline 
$SU(2)_L$ & $\bm{1}$  & $\bm{1}$  & $\bm{1}$  & $\bm{1}$  & $\bm{2}$  & $\bm{1}$  & $\bm{1}$ & $\bm{1}$   \\\hline 
$U(1)_Y$   & $\frac23$ & $-\frac13$ & $-1$ & $0$  & $\frac12$ & $0$  & $0$  & $0$  \\\hline
$U(1)_H$   & $4(1)$ & $-4(-1)$ & $-4(-1)$   & $4(1)$  & $4$  & $-3$  & $-2$  & $1$ \\\hline
$Z_2$ & $-$ & $-$ & $-$ & $+$ & $+$ & $+$ & $+$ & $-$ \\ \hline
\end{tabular}
\caption{ 
Charge assignments of the our fields
under $SU(3)_C\times SU(2)_L\times U(1)_Y\times U(1)_{H} \times Z_2$, where all the SM fields are zero charges under the $U(1)_H$ symmetry and even under the $Z_2$.}
\label{tab:1}
\end{table}

\begin{figure}[t]
\centering
\includegraphics[width=0.60\textwidth]{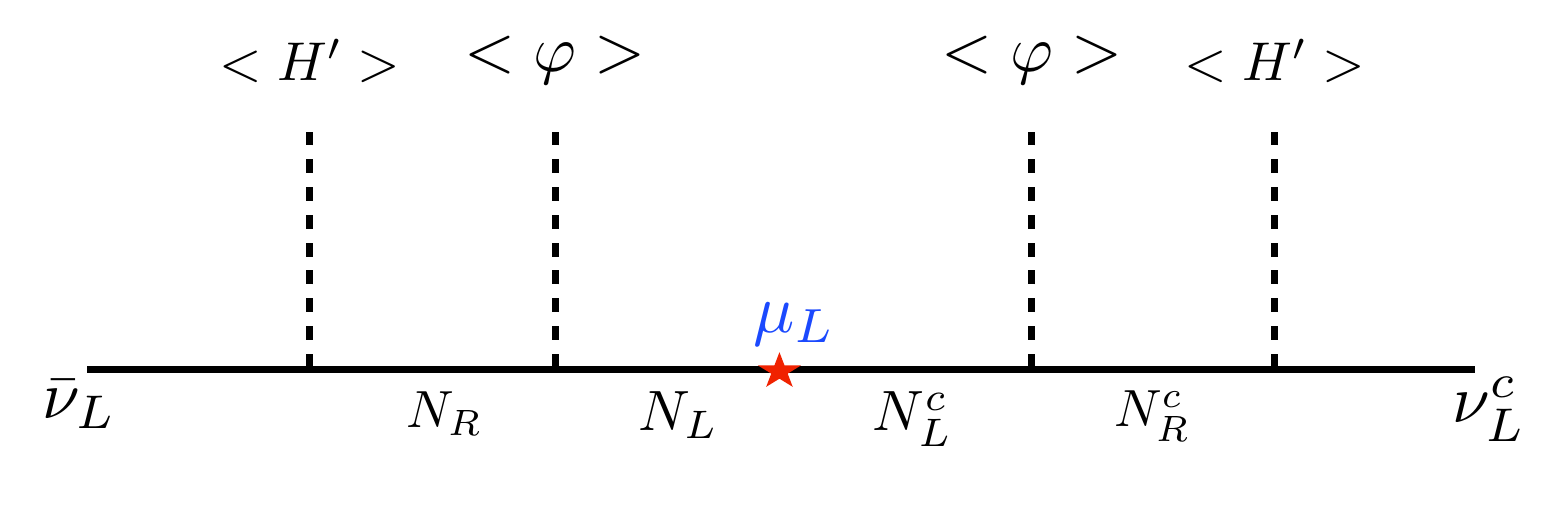} 
\caption{
The masses for active neutrinos are   generated by  inverse seesaw.}
\label{fig:neutrino}
\end{figure}

We will start by  presenting  the particle content in our model. First of all, we introduce three families of right(left)-handed vector-like fermions $U,D,E,N$ which are charged under $U(1)_H$ gauge symmetry; note that actually they are chiral under $U(1)_H$ and become vector-like fermions after its spontaneous symmetry breaking~\cite{Ko:2016ala, Ko:2016wce}. To have gauge anomaly-free for $[U(1)_H]^2[U(1)_Y]$ and  $[U(1)_H][U(1)_Y]^2$, the number of family has to be the same for each fermion, although $[U(1)_H]^3$ and  $[U(1)_H]$ are anomaly free between $U$ and $D$ or $E$ and $N$.\footnote{We can show the non-trivial anomaly free conditions for $[U(1)_H]^2[U(1)_Y]$ and  $[U(1)_H][U(1)_Y]^2$.  For $[U(1)_H]^2[U(1)_Y]$: $n_f \left[3 \cdot \frac{2}{3} (4^2 -1) - 3\cdot \frac{1}{3} (4^2 -1)- (4^2-1)\right] =0$; For $[U(1)_H][U(1)_Y]^2$: $n_f \left[ 3\cdot \left(\frac{2}{3}\right)^2 (4 -1)+ 3\cdot \left(-\frac{1}{3}\right) (4-1) + (-4 +1 )\right] =0$. The  $n_f$ is required  to be the same  for $U, D, E$ so that  the anomaly cancellation is achieved. In this model, we can set $n_f = 3$.} In scalar sector, we add  an isospin doublet boson $H'$ with charge 4 under the $U(1)_H$ symmetry that plays an role in having Dirac mass terms in the neutrino sector after spontaneous electroweak symmetry breaking.  Also we require three isospin singlet bosons $(\varphi,\varphi', \chi)$ with charges $(-3,-2,1)$ under the $U(1)_H$ symmetry, where $\varphi,\varphi'$ have nonzero vacuum expectation values to induce masses for $U,D,E,N$, while $\chi$ is expected to be an inert boson that can be a DM candidate.  Here, we denote that all the SM fields are neutral under $U(1)_H$ symmetry, and each of vacuum expectation value is symbolized by $\langle H^{(')}\rangle\equiv v_{H^{(')}}/\sqrt2$, and $\langle \varphi^{(')}\rangle\equiv v_{\varphi^{(')}}/\sqrt2$, where $H$ is the SM Higgs field.
In addition we introduce $Z_2$ symmetry assigning odd parity to $\{\chi, U, D, E \}$  so that   the stability of $\chi$ is guaranteed as a dark matter (DM) candidate.
The $Z_2$ parity forbids additional interaction terms: $\lambda_0 \chi \varphi^* \varphi^{\prime 2}$, $\lambda_0 \left( H'^{\dagger} H\right) \varphi^* \chi$ and $\mu_0 \chi \varphi \varphi^{\prime *}$, which are permitted by the $U(1)_H$ symmetry but could lead to the decay of $\chi$ into SM particles.
All the new field contents and their charge assignments are summarized in Table~\ref{tab:1}.
The relevant renormalizable Yukawa Lagrangian and  Higgs potential under these symmetries are given by
\begin{align}
-{\cal L}_Y
&=  y_{N_{aa}} \bar L_{L_a} \tilde H' N_{R_a}  +  y_{{N\varphi}_{aa}} \bar N_{L_a} N_{R_a} \varphi
+  y_{{N\varphi'}_{ab}} \bar N^C_{L_a} N_{L_b} \varphi' 
  +  y_{{U\varphi}_{aa}} \bar U_{R_a} U_{L_a} \varphi^* 
 + y_{{D\varphi}_{aa}} \bar D_{R_a} D_{L_a} \varphi\nn\\
&
+  y_{{E\varphi}_{aa}} \bar E_{R_a} E_{L_a} \varphi  
+ (y_{u\chi})_{ia}\bar u_{R_i} U_{L_a} \chi^* 
+ (y_{d\chi})_{ia}\bar d_{R_i} D_{L_a} \chi 
+(y_{e\chi})_{ia} \bar e_{R_i} E_{L_a} \chi
+ {\rm h.c.}, \label{Eq:yuk} 
\end{align}

\begin{align}
V&= \sum_{\phi}^{H,H',\varphi,\varphi',\chi}\left[\mu^2_\phi \phi^\dag \phi +  \lambda_\phi |\phi^\dag \phi|^2 \right]
+ \frac12 \sum_{\phi\neq \phi'}^{H,H',\varphi,\varphi',\chi} \lambda_{\phi\phi'} |\phi|^2 |\phi'|^2
+ \lambda'_{HH'} (H^\dagger H')(H'^\dagger H)
\nn\\&
+\left[   \lambda_0(H^\dag H') \varphi'^2 - \mu \chi \chi \varphi'   + {\rm h.c.}\right], \label{Eq:pot}
\end{align}
where $\tilde H =  i\sigma_2H^*$, $\lambda^{(')}_{\phi\phi'}=\lambda^{(')}_{\phi'\phi}$, and the upper indices $(a,b,i)=1, 2, 3$ are the number of families. 
All the  Yukawa couplings in Eq.~(\ref{Eq:yuk}) are assumed to be diagonal  except for  $y_{N \varphi'}$. Thus in this model,  the mixing of active neutrinos are induced via $y_{N \varphi'}$ and as illustrated by Figure~\ref{fig:neutrino}, the neutrino mass is generated by the inverse seesaw. With the outline of particle content and   Lagrangian, we are going to  present the detail for each sector.
 %

\subsection{Scalar sector}
We will first  focus on the scalar spectra  by demanding  the  VEV of $\chi$ to be vanishing. The non-zero VEVs of scalar fields are obtained by the minimum conditions:
\begin{equation}
\frac{\partial V}{\partial v_{H}} = \frac{\partial V}{\partial v_{H'}} =\frac{\partial V}{\partial v_{\varphi}} = \frac{\partial V}{\partial v_{\varphi'}} =0,
\end{equation} 
Since  the   $v_{H'}$  generates a   mass term of $\bar L H' N_R$, it  is natural  to be $\mathcal{O}(1)$ GeV. On the other hand  the $v_\varphi$ gives a Dirac mass  to extra fermions $N_{L(R)}$  which is expected to be in TeV scale (refer to Section~\ref{sec:neutrino} for more detail). Thus we can impose a VEV hierarchy of $v_{H'} \ll v_H < v_{\varphi'} \ll v_{\varphi}$ in order  to realise  the inverse seesaw mechanism. In this limit,  we will approximately obtain the  expressions:
\begin{align}
& v_H \simeq \sqrt{\frac{2 \left( \lambda _{H \varphi '} \mu _{\varphi'}^2 -2 \lambda _{\varphi'} \mu _{H}^2\right)}{4 \lambda _H \lambda _{\varphi '} -\lambda_{H \varphi'}^2}}
, \quad ~~ v_{\varphi'} \simeq \sqrt{\frac{2 \left( \lambda _{H \varphi '} \mu _H^2  -2 \lambda _H \mu _{\varphi '}^2\right)}{4 \lambda _H \lambda _{\varphi '} -\lambda_{H \varphi'}^2}} \nn \\
& v_{H^\prime} \simeq \frac{- \lambda_0 v_H v_{\varphi'}^2 }{2 \mu_{H'}^2 + (\lambda_{H H'} + \lambda'_{H H'})v_H^2}, \quad v_\varphi \simeq \sqrt{\frac{- \mu_\varphi^2}{\lambda_\varphi}}
\end{align} 
assuming $\{\lambda_{H \varphi},  \lambda_{H' \varphi}, \lambda_{H' \varphi'}, \lambda_{\varphi \varphi'} \} \ll 1$. Since we prefer a notable mixing between $H$ and $\varphi'$ to induce  DM-nucleon scattering, the coupling  $\lambda_{H\varphi'}$  is only slightly less than $\lambda_{H}, \lambda_{\varphi'}$.  And we require $\{\left(4 \lambda _H \lambda _{\varphi '} -\lambda_{H \varphi'}^2\right),  \left( \lambda _{H \varphi '} \mu _{\varphi'}^2 -2 \lambda _{\varphi'} \mu _{H}^2\right),  \left( \lambda _{H \varphi '} \mu _H^2  -2 \lambda _H \mu _{\varphi '}^2\right)\} >0$,  and  $2 \mu_{H'}^2  + (\lambda_{H H'} + \lambda'_{HH'}) v_H^2 >0 $ plus $\{-\mu_\varphi^2,  -\lambda_0 \} > 0$ to make all VEVs positive. The smallness of $v_{H'}$ can be achieved by requiring  $\lambda_0$  to be  negligible, so that the $v = \sqrt{v_{H}^2 + v_{H'}^2}  = 246$ GeV is mainly determined by the $v_H$. The two Higgs doublet fields are parameterized to be:
\begin{align}
&H =\left[\begin{array}{c}
w^+\\
\frac{v_H + h +i \eta}{\sqrt2}
\end{array}\right],\quad 
H' =\left[\begin{array}{c}
w'^+\\
\frac{v_{H'} + h' + i \eta'}{\sqrt2}
\end{array}\right],
\end{align}
where one massless combination of  the charged scalars $w^+, w'^+$  is absorbed by the SM  gauge boson $W^\pm$, and one degree of freedom composed by the CP-odd scalars $\eta$ and $\eta'$ is  eaten by the neutral SM gauge boson $Z$. 
In the case of $v_{H'} \ll v_H$ we can approximate 
\begin{equation}
w^+ \simeq G^+, \quad \eta \simeq G_Z, \quad w'^+ \simeq H^+, 
\end{equation}
Here $G^+$ and $G_Z$ indicate Nambu-Goldstone boson and $H^+$ is remaining physical charged Higgs boson, same as the two-Higgs doublet models.

In the symmetry breaking phase, we also have massless
 Nambu-Goldstone (NG) boson absorbed by $Z'$ and physical Goldstone boson from singlet scalar fields $\varphi$ and $\varphi'$. 
To discuss these massless bosons we first write $\varphi$ and $\varphi'$  by:
\begin{equation}
\varphi=
\frac{v_\varphi + \varphi_{R}}{\sqrt2}{e^{i \frac{ \alpha}{v_\varphi }}},  \quad 
\varphi'=
\frac{v_{\varphi'} + \varphi'_{R}}{\sqrt2} {e^{i \frac{ \alpha'}{v_{\varphi'} } }}.
\label{eq:component}
\end{equation}
While the NG boson $\alpha_{NG}$ and physical Goldstone boson $\alpha_G$ should be recasted  in terms of linear combination of $\alpha$ and $\alpha'$, with the mixing angle  determined by relative sizes of VEVs.  We  therefore  obtain the  expression for the NG boson   and physical Goldstone boson:
\begin{align}
  \alpha_{NG} & = c_X \alpha + s_X \alpha', ~ \quad ~ \alpha_G = - s_X \alpha + c_X \alpha' , \label{eq:GandNG1} \\
 c_X \equiv \cos X &= \frac{3 v_\varphi}{\sqrt{9 v_\varphi^2 + 4 v_{\varphi'}^2 }}, \quad s_X \equiv \sin X = \frac{2 v_{\varphi'}}{\sqrt{9 v_\varphi^2 + 4 v_{\varphi'}^2 }}. \label{eq:GandNG2}
\end{align} 
and we can  simply write $\alpha_{NG} \simeq \alpha$ and $\alpha_G \simeq \alpha'$ for the sake  of $v_{\varphi'} \ll v_{\varphi}$.

For neutral CP-even scalar bosons, we obtain the mass matrix in  basis of $(h, h', \varphi_R, \varphi'_R)$ as follows:
{\footnotesize{
\begin{equation}
M^2_\phi \simeq \left( \begin{array}{cccc} 2 \lambda_H v_H^2 & \left(\lambda_{HH'} + \lambda'_{HH'} \right) v_H v_{H'} & \lambda_{H \varphi } v_H v_\varphi & \lambda_{H\varphi'} v_H v_{\varphi'} + \lambda_0 v_{H'} v_{\varphi'}\\
\left(\lambda_{HH'} + \lambda'_{HH'} \right) v_H v_{H'}  & -\frac{\lambda_0}{2} \frac{v_H}{v_{H'}} v_{\varphi'}^2 & \lambda_{H' \varphi} v_{H'} v_\varphi & \lambda_{H' \varphi'} v_{H'} v_{\varphi'} +\lambda_0 v_{H} v_{\varphi'}\\
\lambda_{H \varphi } v_H v_\varphi &  \lambda_{H' \varphi} v_{H'} v_\varphi & 2 \lambda_\varphi v^2_\varphi & \lambda_{\varphi \varphi'} v_\varphi v_{\varphi'} \\
 \lambda_{H\varphi'} v_H v_{\varphi'} + \lambda_0 v_{H'} v_{\varphi'} & \lambda_{H' \varphi'} v_{H'} v_{\varphi'} +\lambda_0 v_{H} v_{\varphi'} & \lambda_{\varphi \varphi'} v_\varphi v_{\varphi'} & 2 \lambda_{\varphi'} v^2_{\varphi'}
\end{array} \right),
\end{equation}}}
To induce DM-nucleon scattering, we can assume only two CP-even scalars  $(h, \varphi'_R)$  have sizable mixing. This can be realised  by setting  the corresponding coupling for other mixing to be tiny: $ \{\lambda_{H H'}, \lambda'_{H H'}, \lambda_{H \varphi} \}\sim \mathcal{O}\left(\frac{v_{H'}}{v_H}\right)^2$ and $ \{\lambda_{H' \varphi}, \lambda'_{H \varphi'}, \lambda_{\varphi \varphi'}, \lambda_0 \}\sim \mathcal{O}\left(\frac{v_{H'}}{v_\varphi}\right)^2$. Thus the mass eigenvalues reads:
\begin{align}
& m_{h}^2 \simeq   \lambda_H v_H^2 + \lambda_{\varphi'} v_{\varphi'}^2 - \sqrt{(\lambda_H v_H^2 - \lambda_{\varphi'} v_{\varphi'}^2)^2 + \lambda_{H \varphi'} ^2 v_H^2 v_{\varphi'}^2 }, \label{eq:scalar-mass1} \\
& m_{H_1}^2 \simeq   \lambda_H v_H^2 + \lambda_{\varphi'} v_{\varphi'}^2 + \sqrt{(\lambda_H v_H^2 - \lambda_{\varphi'} v_{\varphi'}^2)^2 + \lambda_{H \varphi'}^2 v_H^2 v_{\varphi'}^2 }, \label{eq:scalar-mass2} \\
& m_{H_2}^2 \simeq -\frac{\lambda_0}{2} \frac{v_H}{v_{H'}} v_{\varphi'}^2, \\
& m_{H_3}^2 \simeq 2 \lambda_\varphi v_\varphi^2, 
\end{align}
and the  mixing among $h$ and $\varphi'_R$ is parameterised as 
\begin{align}
 \left( \begin{array}{c} h \\ \varphi'_R \end{array} \right) & = \left( \begin{array}{cc} \cos \theta_{h} & - \sin \theta_{h} \\ \sin \theta_h & \cos \theta_h \end{array} \right) 
\left( \begin{array}{c} h_{SM} \\ H_1 \end{array} \right), \nonumber \\
 \tan  \theta_h  & =  - \frac{1}{\lambda_{H \varphi'} v_H v_{\varphi'}} \left[\sqrt{(\lambda_{H} v_H^2 - \lambda_{\varphi'} v_{\varphi'}^2)^2 +  \lambda_{H\varphi'}^2 v_H^2 v_{\varphi'}^2}  \right.  \nn \\ &  \left.  -\left(\lambda_{\varphi'} v_{\varphi'}^2 - \lambda_H v_H^2 \right)\right].
\label{eq:scalar-mixing}
\end{align}
Note that if the quartic coupling  $\lambda_{H\varphi'}$ is tuned  to be large enough,  this will  result in a sizable mixing angle. 
In such  case the  Higgs coupling is universally rescaled by a mixing angle and its current bound is $\sin \theta_h \lesssim 0.3$ from the analysis of Higgs precision measurements~\cite{Chpoi:2013wga, Cheung:2015dta}.
In addition, the mixing between those CP-even scalars will cause invisible Higgs decays $h \to \chi_{R(I)} \chi_{R(I)} $ depending on the mass spectrum  as well as $h \to \alpha_G \alpha_G$ via the kinematic term.   Since we plan to  take  $m_{\chi_{R(I)}} > m_h/2$  in our analysis below,  thus only  the process $h \to \alpha_G \alpha_G$ will be considered here.
From the kinetic terms of $\varphi'$ we obtain 
\begin{equation}
\mathcal{L} \supset \frac{1}{2 v_{\varphi'}} \varphi'_R \partial_\mu \alpha_G \partial^\mu \alpha_G = 
\frac{\sin \theta_h}{2 v_{\varphi'}} h_{SM} \partial_\mu \alpha_G \partial^\mu \alpha_G + \frac{\cos \theta_h}{2 v_{\varphi'}} H_1 \partial_\mu \alpha_G \partial^\mu \alpha_G,
\end{equation}
where we applied the scalar mixing in Eq.~(\ref{eq:scalar-mixing}).
The decay width of $h_{SM} \to \alpha_G \alpha_G$ process is given by
\begin{equation}
\Gamma_{h_{SM} \to \alpha_G \alpha_G} = \frac{m_h^3 \sin^2 \theta_h}{256 \pi v_{\varphi'}^2}.
\end{equation}
Then the branching ratio is estimated as 
\begin{equation}
BR(h_{SM} \to \alpha_G \alpha_G) \simeq  0.052 \left( \frac{1000 \ {\rm GeV}}{v_{\varphi'}} \right)^2 \left( \frac{\sin \theta_h}{0.3} \right)^2,
\end{equation}
Thus for  $\Gamma_h = 4.19$ MeV, $v_{\varphi' }> 500$ GeV and $\sin \theta_h \sim 0.1$,  it is safe from the current upper bound $BR_{\rm invisible} < 0.25$~\cite{Aad:2015pla}. For phenomenology interest, we should consider the branching ratio of $H_1$ decay  since it can be produced at the LHC via  scalar mixing. We find out that depending on the $\mu$ parameter,  the $H_1$  mainly decays into $\chi_{R(I)} \chi_{R(I)}$, $\alpha_G \alpha_G$ plus SM particles. While  the $H_1$ decay  involving either $\varphi_R$ or  $H'$ are subdominant for  $\lambda_{\varphi \varphi'}, \lambda_{H'\varphi'} \ll 1$, so is  its decay into SM Higgs pair.  The last point can be  illustrated  by an explicit calculation. The interactions inducing $H_1 \to h_{SM}h_{SM}$  are  expressed by:
\begin{eqnarray}
\mathcal{L} & \supset & \bigg(3 \lambda_{\varphi'} v_{\varphi'} \sin^2 \theta_h \cos \theta_h + \frac{1}{2} \lambda_{H\varphi'} v_\varphi' (-2 \sin \theta_h \cos^2 \theta_h + \cos^3 \theta_h ) \nn \\ &-&  3  \lambda_{H} v_{H}  \sin \theta_h  \cos^2 \theta_h  +  \frac{1}{2} \lambda_{H\varphi'} v_H (2 \sin \theta_h \cos^2 \theta_h -\sin^3 \theta_h )  \bigg) H_1 h_{SM}^2 
\end{eqnarray}
where the $H_1 h_{SM}^2$ coupling depends on $\lambda_{\varphi'}, \lambda_H, \lambda_{H\varphi'}$ if we fix $v_H \simeq 246$ GeV and $v_{\varphi'} = 1000$ GeV. Three couplings will be solved by the conditions of $m_h = 125$ GeV, $m_{H_1} = 500$ GeV and a specific value of $\sin \theta_h$ using Eq.(\ref{eq:scalar-mass1} -\ref{eq:scalar-mass2}) and Eq.(\ref{eq:scalar-mixing}).  This gives  $\Gamma_{H_1 \to h_{SM} h_{SM}} =  \frac{C_{Hhh}^2}{32 \pi m_{H_1}} \sqrt{1- \frac{4m_{h}^2}{m_{H_1}^2}} < 6 \cdot 10^{-6}$ for  $\sin \theta_h < 0.3$.
The partial decay widths for the major $H_1$ decay channels  are written as: 
\begin{align}
& \Gamma_{H_1 \to \alpha_G \alpha_G} = \frac{m_{H_1}^3 \cos^2 \theta_h}{256 \pi v_{\varphi'}^2}  \\
& \Gamma_{H_1 \to X X} =  \frac{\mu^2 \cos^2 \theta_h}{16 \pi m_{H_1}} \sqrt{1 - \frac{4 m_{X}^2}{m_{H_1}^2}}.
\\
&  \Gamma_{H_1 \to \bar{t}~ t} = \frac{3 G_{F} m_{H_1} m_{t}^2 \sin^2 \theta_h }{4 \sqrt{2} \pi } \left(1-\frac{4
   m_t^2}{m_{H_1}^2}\right)^{3/2}   \\
  &   \Gamma_{H_1 \to W W } = \frac{G_{F} m_{H_1}^3 \sin^2 \theta_h }{8 \sqrt{2} \pi } \sqrt{1-\frac{4 m_{W}^2}{m_{H_1}^2}} \left(\frac{12
   m_{W}^4}{m_{H_1}^4}-\frac{4 m_{W}^2}{m_{H_1}^2}+1\right)  \\
     &   \Gamma_{H_1 \to Z Z } = \frac{G_{F} m_{H_1}^3 \sin^2 \theta_h }{16 \sqrt{2} \pi } \sqrt{1-\frac{4 m_{Z}^2}{m_{H_1}^2}} \left(\frac{12
   m_{Z}^4}{m_{H_1}^4}-\frac{4 m_{Z}^2}{m_{H_1}^2}+1\right)
\end{align}
Note that  $\Gamma_{H_1 \to \chi_I \chi_I} = \Gamma_{H_1 \to X X }$ in case of  $m_{\chi_I} = m_{X}$, which should be summed up into the total width.  
\begin{figure}[t]
\centering
\includegraphics[width=0.42\textwidth]{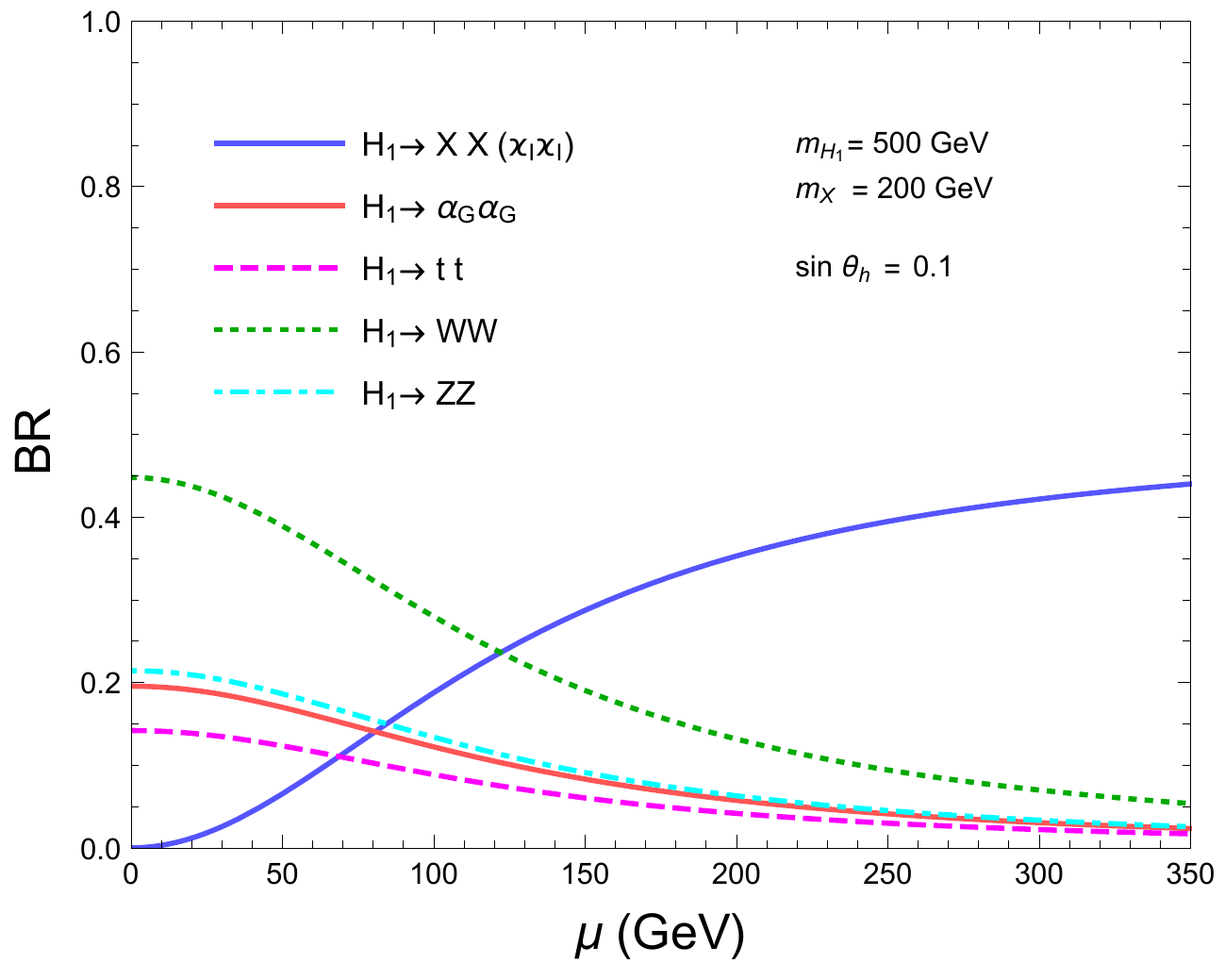} \qquad 
\includegraphics[width=0.42\textwidth]{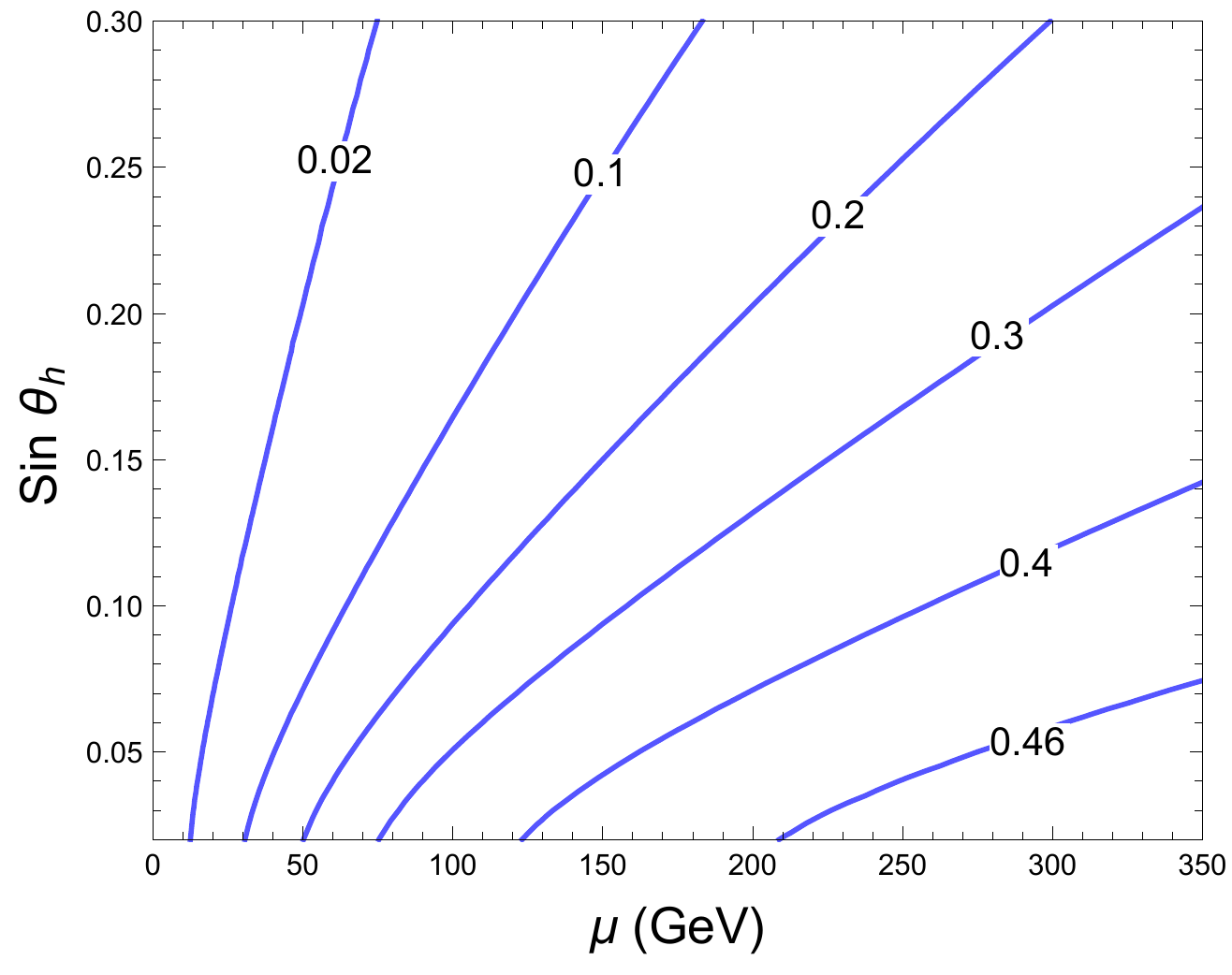}
\caption{
Left plot: The branching ratios  of $H_1$ into each channel are visualised  as  the functions of $\mu$ with $m_{H_1} = 500$ GeV, $m_X = 200$ GeV, and $\sin \theta =0.1 $. Right plot: The branching ration of $BR(H_1 \to X X)$  as contours in the plane of $(\mu, \sin \theta)$ with $m_{H_1} = 500$ GeV and  $m_X = 200$ GeV.
}
\label{fig:BRH1}
\end{figure}  
In Fig.~\ref{fig:BRH1} we present the dependence of $BR(H_1 \to \rm{anything})$ on a single variable $\mu$ and the contour of $BR(H_1 \to XX)$ in the plane of $(\mu, \sin \theta_h)$, with other parameters indicated in the caption. The plots show that in the low $\mu$ region,  $H_1$ mainly decays into $\alpha_G \alpha_G$ and SM particles $\bar t t$, $WW$ and $ZZ$ regardless of the mixing angle.  While near the corner of   large $\mu$ and small  $\sin \theta_h$,  the dominant decay of $H_1$ is into DM and its partners $XX + \chi_{I} \chi_{I}$.  For   $\mu \sim 170$ GeV and  $\sin \theta \sim 0.05$,  we roughly obtain  $BR_{H_1 \to XX} \simeq 0.4 $.

The $Z_2$ odd scalar $\chi$ is written as $\chi = (\chi_R + i \chi_I)/\sqrt{2}$.
The masses for each component are given by
\begin{align}
m_{\chi_R}^2 = \mu_\chi^2 + \frac{1}{2} (\lambda_{H \chi} v_H^2 + \lambda_{H' \chi} v_{H'}^2 + \lambda_{\varphi \chi} v_{\varphi}^2 + \lambda_{\varphi' \chi} v_{\varphi'}^2) - \sqrt{2} \mu v_{\varphi'} \\
m_{\chi_I}^2 = \mu_\chi^2 + \frac{1}{2} (\lambda_{H \chi} v_H^2 + \lambda_{H' \chi} v_{H'}^2 + \lambda_{\varphi \chi} v_{\varphi}^2 + \lambda_{\varphi' \chi} v_{\varphi'}^2) + \sqrt{2} \mu v_{\varphi'}, 
\end{align} 
where the last term in right-hand side provides the mass difference between $\chi_R$ and $\chi_I$. Depending on the sign of $\mu$ coupling, either the real or the imaginary part of the $\chi$ scalar can be the DM candidate.

\subsection{Neutrino sector} \label{sec:neutrino}
After the spontaneous  symmetry breaking, one  has neutral fermion masses which are defined by $m_D\equiv y_N v_{H'}/\sqrt2$, $M\equiv y_{N\varphi} v_\varphi/\sqrt2$, and  $\mu_L\equiv y_{N\varphi'} v_{\varphi'}/\sqrt2 $.   
Then, the neutral fermion mass term  in the basis of $\left( \nu_L^i , N_R^{C, \, a}, N_L^{a} \right)$, $(i, a)= 1,2,3$,  is given by
\begin{align}
M_N
&= 
\left[\begin{array}{ccc}
0 & m_D^* & 0  \\ 
m_D^\dag & 0 & M^T \\ 
0  & M & \mu_L \\ 
\end{array}\right] 
\end{align}
The active neutrino mass matrix can be approximated as:
\begin{align}
m_\nu\approx m_D^* M^{-1} \mu_L (M^T)^{-1} m_D^\dag,
\end{align}
which  can be directly calculated from  Feynman diagram as well  under  the seesaw limit  of $\mu_L  \lesssim m_D \ll M$ and   assuming that   $\mu_L$($= \mu_L^T$), $M$ to be real.
The neutrino mass  (9$\times$9) matrix  is diagonalized by a unitary matrix $U_{MNS}$, i.e. $D_\nu= U_{MNS}^T m_\nu U_{MNS}$, with $D_\nu =  {\rm diag}(m_1,m_2,m_3)$. 
One of the elegant ways to reproduce the current neutrino oscillation data~\cite{Forero:2014bxa} is to apply the Casas-Ibarra parametrization~\cite{Casas:2001sr}. Without loss of generality,  we find the following relation:
\begin{align}
m_D=U_{MNS} \sqrt{D_\nu} O_{mix}  R_N^{-1}.
\end{align}
where $O_{mix}$ is an arbitrary 3 by 3 orthogonal matrix with complex values,  and $R_N$ is a lower unit triangular~\cite{Nomura:2018ktz}, which can uniquely be decomposed to be $\mu_M  \equiv M^{-1} \mu_L (M^T)^{-1}=R_N R^T_N$, since it is symmetric.  For clarity,  we  provide the explicit form of $R_N$ in term of the  elements of $\mu_M = M^{-1} \mu_L (M^T)^{-1}$~\footnote{The $R_N$ parametrisation  in ref.~\cite{Nomura:2018ktz} is not fully correct.}:
\begin{align}
&
 R_N^{-1}=
\left[\begin{array}{ccc}
\frac1a & 0 & 0  \\ 
-\frac{d}{ab} & \frac{1}{b} & 0 \\ 
\frac{-be+df}{abc}  & \frac{f}{bc} &\frac{1}{c}  \\ 
\end{array}\right]
\end{align}
\begin{align}
& a=\sqrt{\mu_{M, 11}},\quad d=\frac{\mu_{M, 12}}{a},\quad b=\sqrt{\mu_{M, 22}-d^2}, \quad  f=\frac{ d~\mu_{M,13}  - a~ \mu_{M, 23} }{a b} \nn \\
&
e=\frac{\mu_{M, 13}}{a} + 2 \frac{d}{b} f,\quad  c=\sqrt{\mu_{M,33}-\left(e-2 \frac{d}{b} f \right)^2-f^2},
\end{align}
Note  that  the absolute value of all components in $m_D$ should not exceed $\frac{1}{\sqrt{2}}$ GeV with $v_{H'}=1$ GeV, once the perturbative limit for $|y_N| =  |\sqrt2m_D/v_{H'}|$ is taken to be 1.

{\it Non-unitarity}: 
We should mention the possibility of non-unitarity matrix $U'_{MNS}$ due to the mixing related to  heavy fermions.
This is typically parametrized by the form:
\begin{align}
U'_{MNS}\equiv \left(1-\frac12 FF^\dag\right) U_{MNS},
\end{align}
where $F$ is a hermitian matrix determined by each model, $U_{MNS}$ is the three by three unitarity matrix, while $U'_{MNS}$ represents the deviation from the unitarity. Then $F$ is given by~\cite{Das:2017ski, Das:2012ze,  Das:2017nvm} 
\begin{align}
F&= (M^{T})^{-1} m_D^T = (M^{T})^{-1}  (R^{T}_N)^{-1} O_{mix}^T \sqrt{D_\nu} U^T_{MNS},
 \label{eq:f_inv-lin}
\end{align}
The global constraints are found via several experimental results such as {\it the SM $W$ boson mass $M_W$, the effective Weinberg angle $\theta_W$, several ratios of $Z$ boson fermionic decays, invisible decay of $Z$, EW universality, measured CKM, and LFVs}~\cite{Fernandez-Martinez:2016lgt}.
The result can be given by~\cite{Agostinho:2017wfs} 
\begin{align}
|FF^\dag|\le  
\left[\begin{array}{ccc} 
2.5\times 10^{-3} & 2.4\times 10^{-5}  & 2.7\times 10^{-3}  \\
2.4\times 10^{-5}  & 4.0\times 10^{-4}  & 1.2\times 10^{-3}  \\
2.7\times 10^{-3}  & 1.2\times 10^{-3}  & 5.6\times 10^{-3} \\
 \end{array}\right].\label{eq:unit}
\end{align} 
{We can show a benchmark point  satisfying the observed neutrino masses,  three mixing angles,  plus the Dirac CP violation phase~\cite{deSalas:2017kay},~\footnote{We use the best fit value in the case of normal hierarchy, namely $\Delta m_{21}^2= 7.55 \times 10^{-5}~ {\rm eV}^2$, $\Delta m_{31}^2 =2.50\times 10^{-3} ~{\rm eV}^2$, $\sin^2 \theta_{12} = 0.32$, $\sin^2 \theta_{23} =0.547$, $\sin^2 \theta_{13} = 0.0216$ and $\delta = 1.21 \pi$.  The other parameters are fixed to be  $v_{H'}=1$ GeV and $m_{\nu_1}=10^{-13}$ GeV.}  without conflict with the unitarity bound in Eq.~(\ref{eq:unit}):
\begin{align}
& \frac{\mu_L}{\rm GeV}\approx
\left[\begin{array}{ccc} 
2.8\times 10^{-3} & 5.4\times 10^{-8}  & 8.6\times 10^{-4}  \\
5.4\times 10^{-8}  & 2.5\times 10^{-7}  & 1.2\times 10^{-9}  \\
8.6\times 10^{-4}   &1.2\times 10^{-9} & 5.5\times 10^{-4} \\
 \end{array}\right],\
 \frac{M}{\rm GeV}\approx
\left[
\begin{array}{ccc}
 1605 & 0 & 0 \\
 0 & 1711 & 0 \\
 0 & 0 & 2801 \\
\end{array}
\right],\nn\\
  &\frac{m_{D}}{\rm GeV}\approx
\left[\begin{array}{ccc} 
- 0.018 + 0.064  i & 1.5\times10^{-4} - 4.0\times 10^{-4} i  & -0.1+0.026  i \\
-0.38+0.077 i & 5.8\times10^{-4} - 4.4\times 10^{-4}i  & 0.62+0.21 i  \\
 -0.19- 0.082i  &1.1\times10^{-5}+3.4\times10^{-4}i & 0.58+0.18 i \\
 \end{array}\right],\nn\\
 &O_{mix}\approx
\left[
\begin{array}{ccc}
 0.82\, +0.95 i & 1.1\, -0.38 i & 0.62\, -0.55 i \\
 -0.90+0.68 i & 1.0\, +0.60 i & 0.0091\, -0.0052 i \\
 -0.95+0.17 i & -0.19+0.92 i & 1.0 \, +0.34 i \\
\end{array}
\right].
\end{align}
}
For comparison, we comment on  an alternative  non-unitarity parametrisation:  $U'_{MNS} = \left(1-\alpha \right) \tilde{U}_{MNS}$, where $\alpha$ is a lower triangular matrix. Defining $\eta = \frac{F F^\dagger}{2}$,  the translation from the previous one gives   $\alpha_{\beta \beta} \simeq \eta_{\beta \beta}$, and  $\alpha_{\beta \gamma} \simeq 2  \eta_{\beta \gamma}$.   In fact the latter one  imposes a slightly looser  bound  according to  refs. ~\cite{Blennow:2016jkn, Escrihuela:2016ube},  although being more model-independent. In our inverse seesaw,  the  light  neutrino flavors  decompose into  mass eigenstates as  $\nu^{i}_L \simeq (1- \frac{F F^\dagger}{2}) \nu_m^i - \frac{F}{\sqrt{2}} N_1^a - \frac{F}{\sqrt{2}} N_2^a$,  where the unitarity deviation is  same as in  Type-I seesaw. Thus for  $M \sim \mathcal{O} \mbox{(TeV)} \gg m_D $,  two formalisms are equivalent  up to small corrections.

From the Lagrangian  in Eq.~(\ref{Eq:yuk}), we can derive  the masses for  exotic charged fermions $U, D, E$ after scalars gain VEVs, which are denoted as: $M_U =  y_{U\varphi} v_\varphi/\sqrt2$, $M_D= y_{D\varphi} v_\varphi/\sqrt2$, and $M_E = y_{E\varphi} v_\varphi/\sqrt2$. These  parameters are not correlated to the neutrino oscillation data,  but they should be constrained by  DM  relic density and LHC direct bound.

\subsection{Heavy $Z'$ boson}

Here we briefly discuss the Hidden gauge boson in the model where we assume the gauge kinetic mixing between $U(1)_H$ and $U(1)_Y$ is negligibly small. In such a way,   a massive $Z'$ boson  will arise after the  symmetry is broken, whose mass is given by:
\begin{equation}
m_{Z'} \simeq g_H \sqrt{9 v_{\varphi'}^2 + 4 v_{\varphi}^2 + 16 v_{H'}^2},
\end{equation}
where $g_H$ is the $U(1)_H$ gauge coupling. Note that we  have $Z$-$Z'$ mixing since $H'$ is charged  both  under $SU(2)_L \times U(1)_Y$ and $U(1)_H$ symmetries. However the mixing effect is highly suppressed by a factor of $v_{H'}^2/m_{Z'}^2$ if we take $v_{H'} \ll v_{\varphi}, v_{\varphi^\prime}$. Thus the $Z'$ interaction with  SM particles is very small, which makes its detection potential at the LHC Run-II  evadable.


\section{Flavour  and  Dark matter Bounds}

\vspace{- 5 pt}

As we describe  in  the model part, extra scalars and sterile neutrinos are introduced to realise an inverse seesaw, with their interactions governed by the hidden gauge symmetry $U(1)_H$. In particular, the presence of  an inert scalar $\chi$ and  exotic charged fermions gives rise to  the charged LFVs  and flavor-changing   $Z$ decays. These interactions will  induce a  shift in the muon magnetic moment in an expected order  provided the  Yukawa coupling $(y_{e\chi})_{2 a}$, $a =1, 2, 3$ are relatively large. Due to the $Z_2$ parity,  the real part of $\chi$  is stabilised as a DM candidate for $\mu >0$, so that  its  impact on relic density and DM-nucleon scattering would impose constraints as well.  Another interesting aspect is  the DM  production at the LHC, which is  characterised by a pair of charged leptons plus missing transverse energy in this model. For a better illustration we will first focus on the bounds  from  flavour and DM physics  here  and put the discussion of   LHC  phenomenology  in the next section.

 %
\begin{figure}[t]
\centering
\includegraphics[width=0.66\textwidth]{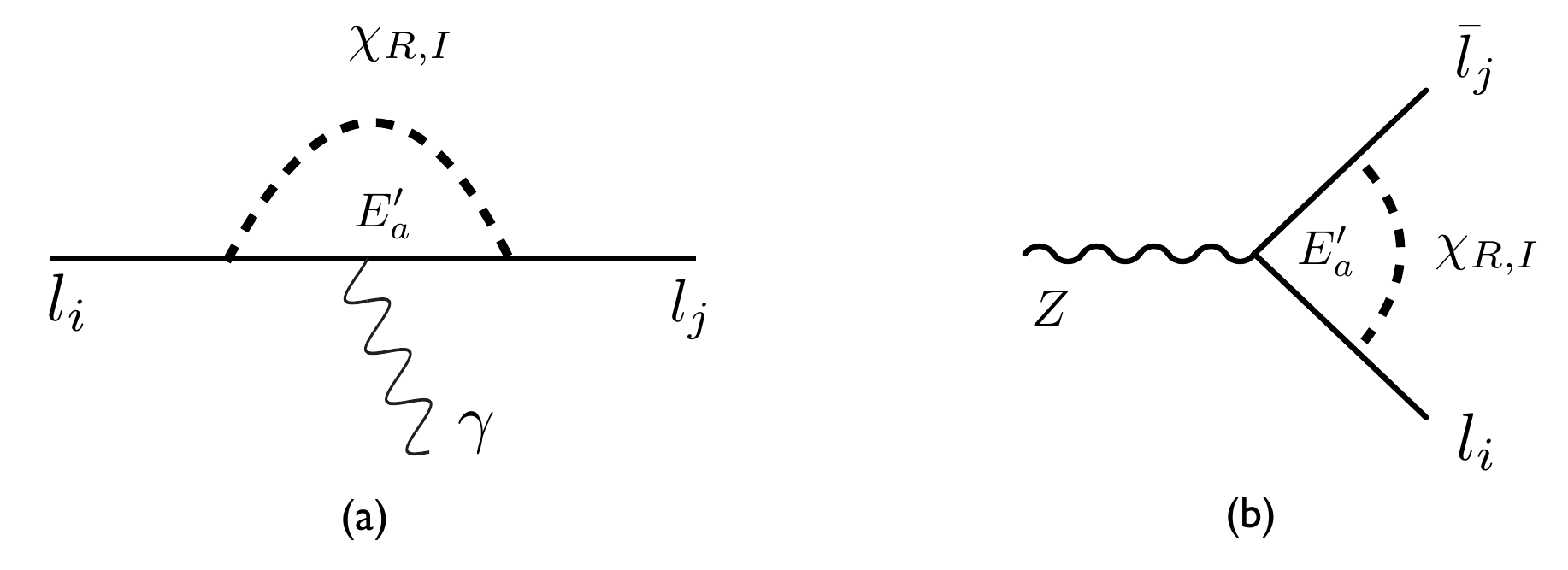} 
\caption{
Feynman diagrams for charged Lepton Flavor Violation decays  and Flavor Changing  leptonic Z decays. Note that for Z decays, in addition to the vertex correction,  the wave function renormalisation should be included  to remove the UV divergence.}
\label{fig:LFV}
\end{figure}  
\subsection{\bf Lepton flavor violations(LFVs)} 
The charged LFV decay can be generated with the mediation of a neutral scalar at the one-loop level. For an inert neutral scalar with no mixing, exotic charged fermions are necessary to present  assuming  the invariance under an extra symmetry such as the $U(1)_H$.  In our model,  the LFV decays  can arise from the Yukawa term $(y_{e\chi})_{ia} \bar{e}_{R, i} E_{L,a } \chi$ as illustrated in  Figure~{\ref{fig:LFV}}(a).  The  branching ratio  ${\rm BR}(\ell_i\to\ell_j\gamma)$ is given by:
\begin{align}
& {\rm BR}(\ell_i\to\ell_j\gamma)= \frac{48\pi^3\alpha_{\rm em} C_{ij} }{{\rm G_F^2} m_{\ell_i}^2}\left(|a_{R_{ij}}|^2+|a_{L_{ij}}|^2\right),\\
 a_{R_{ij}} &\approx
 - \frac{m_{\ell_j}}{2 \, (4\pi)^2}
\sum_{A=R,I}\sum_{a=1,2,3} (y_{e\chi})_{ja}(y^\dag_{e\chi})_{ai} \int[dx]_3 \frac{ x y  }{x M_{\chi_{A}}^2+(1-x) M_{E_a}^2},
\\
\quad 
a_{L_{ij}} &\approx
 - \frac{m_{\ell_i}}{2 \, (4\pi)^2}
\sum_{A=R,I}\sum_{a=1,2,3} (y_{e\chi})_{ja}(y^\dag_{e\chi})_{ai}  \int[dx]_3 \frac{ x y }{x M_{\chi_A}^2+(1-x) M_{E_a}^2},
\end{align}
where $\int[dx]_3\equiv \int_0^1dx\int_0^{1-x}dy$, ${\rm G_F}\approx 1.17\times10^{-5}$[GeV]$^{-2}$ is the Fermi constant, $\alpha_{\rm em}\approx1/137$ is the fine structure constant, $C_{21}\approx1$, $C_{31}\approx 0.1784$, and $C_{32}\approx 0.1736$.
Experimental upper bounds are respectively given by ${\rm BR}(\mu\to e\gamma)\lesssim 4.2\times10^{-13}$, ${\rm BR}(\tau\to e\gamma)\lesssim 3.3\times10^{-8}$, and ${\rm BR}(\tau\to \mu\gamma)\lesssim 4.4\times10^{-8}$~\cite{TheMEG:2016wtm, Adam:2013mnn, Aubert:2009ag}.
\\
{\it New contribution to the muon anomalous magnetic moment} (muon $g-2$: $\Delta a_\mu$)  arises from the same term as in LFVs,
and its  analytic formula reads:\footnote{For a comprehensive review on new physics models for the muon $g-2$ anomaly as well as lepton flavour violation, please see Ref.~\cite{Lindner:2016bgg}.}
\begin{align}
&\Delta a_\mu =- m_\mu [a_R+a_L]_{22}.
\label{eq:G2-ZP}
\end{align}
To explain the current 3.3$\sigma$ deviation~\cite{Hagiwara:2011af}
\begin{align}
\Delta a_\mu = (26.1\pm8.0)\times10^{-10} ~.
\label{eq:damu}
\end{align}
Note  that in case of a large $h-\varphi$ mixing   (by tuning $\lambda_{H\varphi} \sim \lambda_{H \varphi'}$), we  would have  the Barr-Zee type diagrams which contribute to the muon $g-2$. However they only give a small contribution due to   two-loop suppression and  necessity to satisfy  $H\to 2\gamma$ constraint~\cite{Chiang:2017tai}.

\subsection{\bf Flavor-Changing Leptonic $Z$ Boson Decays}\label{subsec:Zll}
As a complementary constraint, we  include  the bound from  the decays of the $Z$ boson into two charged leptons of different flavors  at the one-loop level.
\footnote{Although the quark pairs are also induced from the $y_{u\chi}$ and $y_{d\chi}$, we do not consider them because their experimental bounds are not so stringent. } 
Since we are mainly interested in  the parameter region that  can  achieve a sizeable muon $g-2$,   the flavor-changing  $Z$ decay widths are expected to get non-trivial  contribution from  an ${\cal O}(1)$ Yukawa coupling $(y_{e\chi})_{2 2}$. The relevant form factor  is  obtained  through  the vertex  and wave-function renormalisation  depicted in  Figure~\ref{fig:LFV}(b),  with the analytic expression  calculated to be~\cite{Chiang:2017tai, Fernandez-Martinez:2016lgt}:
\begin{align}
\text{BR}(Z\to\ell^-_i\ell^+_j)
&\approx
\frac{G_F}{12\sqrt2 \pi} \frac{m_Z^3}{(16\pi^2)^2 \Gamma_{Z}^{\rm tot}} 
\left(s_W^2 -\frac12\right)^2
\nn\\& \qquad \times
\left| \sum_{a=1}^{3} \sum_{J = R,I} (y_{e\chi})_{ia} (y_{e\chi}^\dag)_{aj} 
\left[ F_2(E_a,\chi_J)+F_3(E_a,\chi_J) \right] \right|^2 ~,
\label{eq:Zll}
\end{align}
where
\begin{align*}
F_2(a,b) &=\int_0^1dx(1-x)\ln\left[ (1-x)m_a^2 + x m_b^2 \right] ~,
\\
F_3(a,b) &=\int_0^1dx\int_0^{1-x}dy\frac{(xy-1)m_Z^2+(m_a^2-m_b^2)(1-x-y)-\Delta\ln\Delta}{\Delta} ~,
\end{align*} 
with $\Delta\equiv -xy m_Z^2+(x+y)(m_a^2-m_b^2)+m_b^2$ and the total $Z$ decay width $\Gamma_{Z}^{\rm tot} = 2.4952 \pm 0.0023$~GeV. The current upper limit for the lepton flavor-changing $Z$ boson decay branching ratios are published to be~\cite{Patrignani:2016xqp}:
\begin{align}
\begin{split}
  {\rm BR}(Z\to e^\pm\mu^\mp) &< 1.7\times10^{-6} ~,\\
  {\rm BR}(Z\to e^\pm\tau^\mp) &< 9.8\times10^{-6} ~,\\
  {\rm BR}(Z\to \mu^\pm\tau^\mp) &< 1.2\times10^{-5} ~,\label{eq:zmt-exp}
\end{split}
\end{align}
where the upper bounds are quoted at 95 \% CL. After scanning  the parameter space, we found that these constraints are less stringent than  the charged  LFV ones,  and this also applies to  flavor-conserving processes ${\rm BR}(Z\to \ell^\pm\ell^\mp)$ ($\ell=e,\mu,\tau$).

\subsection{\bf Bosonic dark matter candidate}

Fixing  $X= \chi_R$ to be DM, we can first evaluate the relic abundance  by  assuming  the Higgs portal interaction is negligibly small. This hypothesis is quite reasonable  since the $hXX$ coupling is strongly constrained by  the spin independent DM-nucleon scattering as we will discuss later.  The DM annihilations  come from  $XX \to {\bar f} f$ via Yukawa couplings or  $Z'$  boson mediation, although the $Z'$ one is ignorable.   Another  possible channel is   $X  X \to \alpha_G \alpha_G$, where $\alpha_G$ is the physical Goldstone bosons.  To figure out the dominant one,  we can first  examine the couplings. The DM Yukawa interaction  is  directly  read from  Eq~(\ref{Eq:yuk}):
\begin{align}
\frac{ (y_{u\chi})_{ia}}{\sqrt2} \bar u_{R_i} U_{L_a} X +\frac{ (y_{d\chi})_{ia}}{\sqrt2} \bar d_{R_i} D_{L_a} X +\frac{(y_{e\chi})_{ia}}{\sqrt2} \bar e_{R_i} E_{L_a} X  +{\rm H.c.}
~.
\end{align}
 While the DM  interaction with $\alpha_G$ can be derived  from the kinetic term of $\chi$  by a phase rescaling $\chi \to \chi e^{-i \frac{\alpha_G}{2 v_{\varphi'}}}$~\cite{Weinberg:2013kea, Baek:2016wml,Baek:2018wuo}, with  $\alpha' \simeq \alpha_G$  applied:
\begin{align}
&(D_\mu \chi)^\dagger (D^\mu \chi) = \frac{1}{2 v_{\varphi'}} \partial^\mu \alpha_G (\partial_\mu \chi_R \chi_I - \partial_\mu \chi_I \chi_R)
+ g_H Z'^\mu (\partial_\mu \chi_R \chi_I - \partial_\mu \chi_I \chi_R)   \nn \\
&  + \frac{1}{4 v_{\varphi'}^2} \partial_\mu \alpha_G \partial^\mu \alpha_G (\chi_R^2 + \chi_I^2) - \frac{g_H}{v_{\varphi'}} Z'^\mu \partial_\mu \alpha_G (\chi_R^2 + \chi_I^2) + g_H^2 Z'_\mu Z'^\mu (\chi_R^2 + \chi_I^2) 
\end{align}
which  is equivalent to  an  exponential expansion of the term $\chi \chi \varphi'$.  Thus  in the  limit of   $v_H \ll v_{\varphi'}$ and $g_H \ll1$,  plus $\mathcal{O}(1)$ Yukawa couplings favored by the muon $g-2$ anomaly, the majority portion of  required  DM abundance is provided by  the  annihilation induced  by exotic fermions. 
\begin{figure}[t]
\centering
\includegraphics[width=0.5\textwidth]{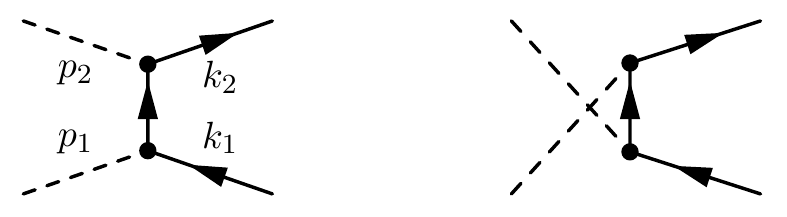} 
\caption{
Feynman diagrams for DM 2-body annihilation into charged SM fermions  $X X \to \bar{f}_{SM} f_{SM}$ via  Yukawa interactions, with  mediators to be  exotic heavy fermions $U, D, E$.}
\label{fig:DManni}
\end{figure}  
We explicitly calculate the  amplitude squared  for  the DM annihilation  process of  $X  X \to \bar{f_i} f_j $   as shown in  Figure~\ref{fig:DManni} to be:
\begin{align}
&  |{\cal{\bar{M}}}_{ij}|^2  = \frac{1}{2} \sum_{a} |(y_{u\chi})_{ia}(y_{u\chi}^\dag)_{aj}|^2  \bigg[ 2 \left(\frac{k_1 \cdot (p_1- k_1)}{(p_1-k_1)^2 -M^2_{U_a}}+ \frac{k_1 \cdot (p_2- k_1)}{(p_2-k_1)^2 -M^2_{U_a}} \right) \nn \\ & \phantom{xxxxxxxxxxx}  \left(\frac{k_2 \cdot (p_1- k_1)}{(p_1-k_1)^2 -M^2_{U_a}}  +\frac{k_2 \cdot (p_2- k_1)}{(p_2-k_1)^2 -M^2_{U_a}} \right)  -   k_1 \cdot k_2  \nn \\ & \phantom{xxxxxxxxxxx}   \left( \frac{(p_1 - k_1)^\mu }{(p_1 - k_1)^2 -M^2_{U_a}} +  \frac{(p_2 - k_1)^\mu }{(p_2 - k_1)^2 -M^2_{U_a}}\right)^2  \bigg] \label{DManni}
\end{align}
with $p_{1,2}$ and $k_{1,2}$ denoting the four momenta of  DM and  SM fermions.  Thus the velocity weighted cross section crucial for the relic density is determined by:
\begin{align}
& \sigma  v =\sum_{i,j} \frac{k_{ij}} {32 \pi^2 s}   \int  d\Omega  |{\cal{\bar{M}}}_{ij}|^2 \,, \rm{with} \quad  s = (p_1 + p_2)^2 \\
&  k_{ij} = \left [1-\frac{(m_{f_i} + m_{f_j})^2}{s} \right]^{1/2} \left[1-\frac{(m_{f_i}-m_{f_j})^2}{s} \right]^{1/2}    
 \end{align}
where  the indices $i, j $ sum over all the SM leptons and quarks. In our case, only  the $\sigma (X X \to \bar t \, t )$ is corrected by a phase space factor of $k_{ij} = \sqrt{1- 4 m_{t}^2/s}$, for other channels   $k_{ij} \simeq 1$ is used in the chiral limit of $m_f/M_X \to 0$.  In  powers of the relative velocity $v_{\rm rel}$ , we get an expansion  $\sigma v  \simeq a_{\rm eff} + b_{\rm eff} v_{\rm rel}^2 + d_{\rm eff} v_{\rm rel}^4$, where $a_{\rm eff}$, $b_{\rm eff}$ and $d_{\rm eff}$ are  s-wave, p-wave and  d-wave  coefficients respectively. Defining  $k_t =  \sqrt{1-  m_{t}^2/m_X^2}$,  the coefficients read:
\begin{align}
& & a_{\rm eff}
\approx 
\frac{3 k_t \, m_t^2}{16 \pi}
 \frac{|(y_{u\chi})_{33}(y_{u\chi}^\dag)_{33}|^2 }{(M_X^2+M^2_{U_a})^2},  \quad b_{\rm eff}  \approx 
-\frac{ k_t \, m_t^2}{8 \pi}  |(y_{u\chi})_{33}(y_{u\chi}^\dag)_{33}|^2 \frac{M_X^2 (M_X^2+ 2 M^2_{U_a})}{(M_X^2+ 2 M^2_{U_a})^4} \,,
\nn \\ 
& &   d_{\rm eff}
\approx 
\frac{ M_X^6}{80 \pi} \sum_{a,i,j, k} 
\left[
\frac{|(y_{u\chi})_{ka}(y_{u\chi}^\dag)_{ak}|^2 }{(M_X^2+M^2_{U_a})^4}+
\frac{|(y_{d\chi})_{ia}(y_{d\chi}^\dag)_{ai}|^2 }{(M_X^2+M^2_{D_a})^4}+
\frac13 \frac{|(y_{e\chi})_{ia}(y_{e\chi}^\dag)_{aj}|^2 }{(M_X^2+M^2_{E_a})^4}
\right]
\label{eq:anni-rl} .
\end{align}
with  $ a, i,j = 1,2,3$ and  $k = 1, 2$ without counting top quark for d-wave, since only  for  $m_f/M_X \to 0$, the $ |{\cal{\bar{M}}}_{ij}|^2$ behaves like  $v_{\rm rel}^4$.  Hence for the  top quark, $s$- and $p$-waves are the leading terms,  but  for those light fermions,  $\sigma_{ij} v$ is $d$-wave dominant.  We  assume that $y_{u\chi}$ and $y_{d\chi}$ to be diagonal to avoid  the constraint from quark sector. For the light fermions, we will take into account the contribution of internal Bremsstrahlung~{\cite{Giacchino:2013bta, Toma:2013bka}}.
\begin{align}
& a_{\rm VIB}=
\frac{3 \alpha_{\mathrm{em}}}{32\pi^2 M_X^2} \sum_{ a, i, j, k } \bigg[ Q_{U_a}^2 |(y_{u\chi})_{k a}(y_{u\chi}^\dag)_{a k}|^2  F \left( \frac{M_{U_a}^2}{M_X^2} \right)   +  Q_{D_a}^2 |(y_{d \chi})_{i a}(y_{d\chi}^\dag)_{a i}|^2  F \left( \frac{M_{D_a}^2}{M_X^2} \right) 
\nn  \\  &  \phantom{xxxxxxxxxxxxxx}    +  \frac{1}{3} Q_{E_a}^2 |(y_{d \chi})_{i a}(y_{d\chi}^\dag)_{a i}|^2  F \left( \frac{M_{E_a}^2}{M_X^2} \right) \bigg]\,,
 \\
&  \phantom{xxxxxx}  F(r) = \left(r+1\right)\left[\frac{\pi^2}{6}-\log^2\left(\frac{r+1}{2 r}\right)
-2\mathrm{Li}_2\left(\frac{r+1}{2 r}\right)\right]  \nn \\
&   \phantom{xxxxxxxxxx} + \frac{4 r+3}{r+1}+\frac{4 r^2-3 r-1}{2 r}\log\left(\frac{r-1}{r+1}
\right) \,.
\end{align}
Due to the fact $a_{\rm VIB} \sim d_{\rm eff} \, v_{\rm ref}^4$ for $r \to 1$, the annihilation cross section is enhanced by a $\mathcal{O}(1)$ boost factor.  The resulting relic density is found to be:
\begin{align}
&\Omega h^2\approx 
\frac{1.07\times 10^9 x_f}{\sqrt{g_*(x_f)} M_{\rm PL} \left[ \left(a_{\rm eff} + + 3 \, b_{\rm eff}/x_f \right) \theta (M_X -m_t) + a_{\rm VIB} + 20 \, d_{\rm eff}/x_f^2\right]} \label{eq:relic-rl} 
\end{align}
\begin{figure}[t]
\centering
\includegraphics[width=0.45\textwidth]{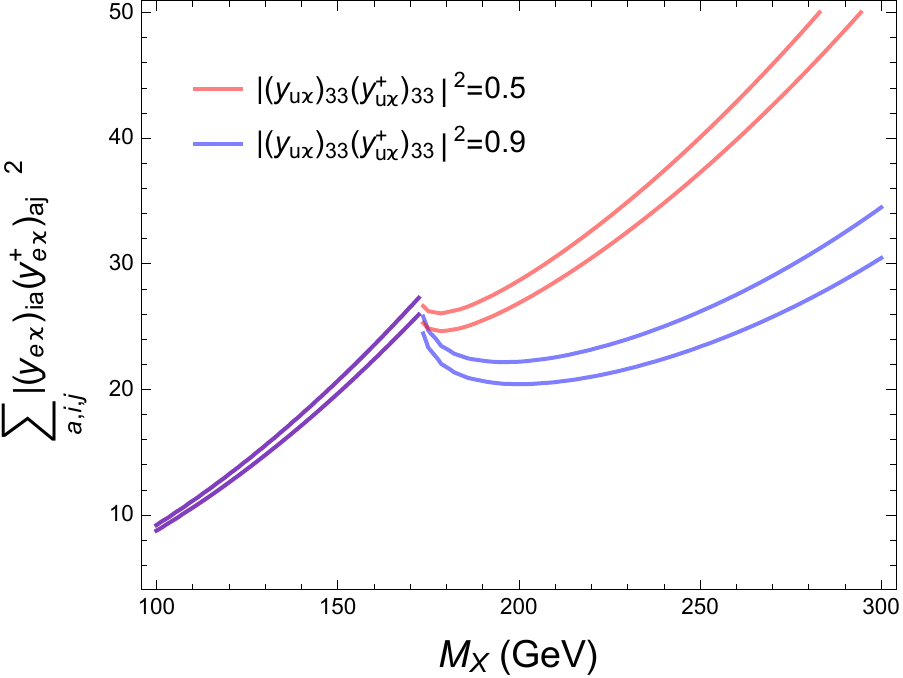} \qquad
\includegraphics[width=0.43\textwidth]{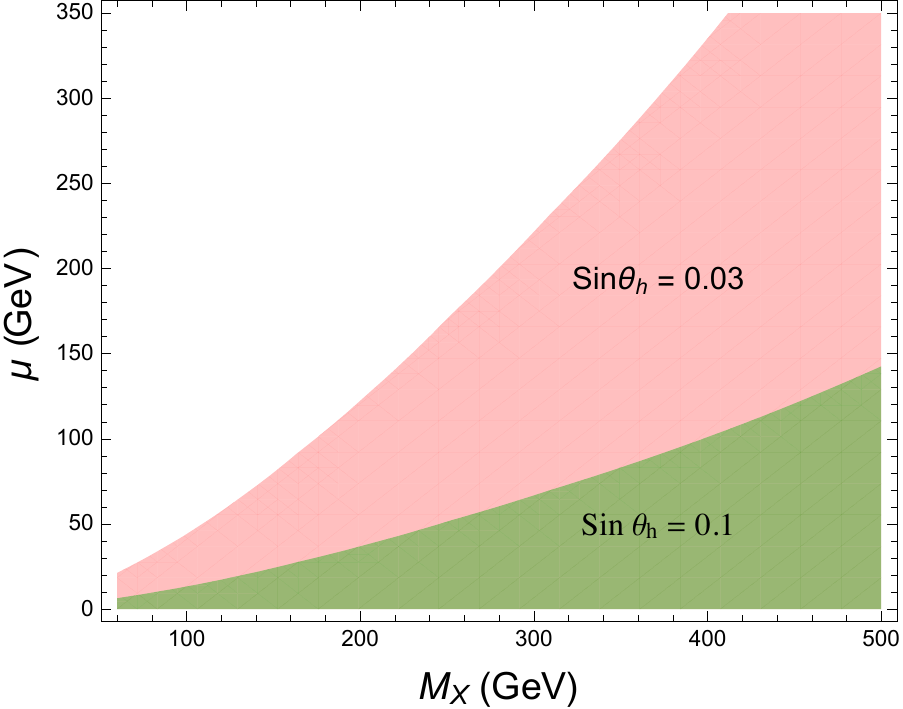}
\caption{
Left plot:  the contours  delimit  the  DM-lepton couplings, which  satisfy the observed relic density $0.12 \pm 0.003$ under the assumption of  $M_U = M_D = 1.0$ TeV, $M_{E} = 1.2 M_X$  and universal DM Yukawa couplings for SM quarks.  Right plot:  the allowed region  for $(M_X, \mu)$  is recasted from the $\sigma_{SI}$ bound in ~\cite{Aprile:2018dbl},  with the  red band for  $\sin \theta =0.03$ and the green band for  $\sin \theta = 0.1$.}
\label{fig:relic&direct}
\end{figure}  
with $ \theta (M_X -m_t) =1$ for $M_X > m_t$, otherwise zero.  Here $g_*(x_f\approx 25)\approx100$ counts the degrees of freedom for relativistic particles, and $M_{\rm PL}\approx 1.22\times 10^{19}$~GeV is the Planck mass. The present relic density is $0.12 \pm 0.003$ at the 3$\sigma$ confidence level (CL)~\cite{Aghanim:2018eyx}. In the left plot of Figure~\ref{fig:relic&direct}, as an estimation, we investigate the sole impact of relic density on DM couplings with SM fermions, where the region between the two lines of same color is allowed.   We take universal Yukawa couplings $y_{u \chi} = y_{d \chi}$  for exotic quarks with  a degenerate mass  $M_{U} = M_{D} = 1.0$ TeV.  While for  all exotic leptons,  we set  their  masses to correlate with the DM mass $M_{E} = 1.2 M_X$,  such that due to  $ r \simeq 1.4$ close to $1.0$, the enhancement for $\langle \sigma v_{\rm rel}\rangle$ from the internal Bremsstrahlung  effect is not negligible. The  plot shows that the annihilation process $X X \to \bar t t$  starts   to   effectuate   beyond the threshold  of  $M_X =  m_t$.  For $M_X = 200$ GeV and  $|(y_{u \chi})_{33} (y^\dag_{u \chi})_{33}|^2 = 0.9$,   the coupling sum  of   $\sum_{a, i, j} |(y_{e \chi})_{i a}(y_{e\chi}^\dag)_{a i}|^2 \sim 20 $  is required   to obtain the correct relic density.  But in  case of a  smaller $|(y_{u \chi})_{33} (y^\dag_{u \chi})_{33}|^2 = 0.5$,   $~\sum_{a, i, j} |(y_{e \chi})_{i a}(y_{e\chi}^\dag)_{a i}|^2  \sim 27$ is expected   for  compensating the reduced   $s$ and $p$ waves contribution from the top quark. A more comprehensive analysis will be explored in the next section, where the lepton flavour bounds are fully included.

DM Direct detection  measures the nucleon recoil  energy for the DM-nucleon scattering in underground experiments. Those searches impose bound for  $(M_X, \mu)$ and $\sin \theta_h$,  so that  the DM production via $H_1$ decay at the LHC  will be discussed afterwards.  The DM-Nucleon scattering is induced via $h_{SM}$ and $H_1$ exchange where the relevant interactions are 
\begin{equation}
\mathcal{L} \supset  \mu \chi_R \chi_R (h_{SM} \sin \theta_h + H_1 \cos \theta_h) + f_N \bar N N (\cos \theta_h h_{SM} - \sin \theta_h H_1)
\end{equation} 
where $N(=p, n)$ denote nucleon field and $f_N = \frac{2}{9} + \frac{7}{9} \sum_{q=u,d,s} f_q$ is the effective coupling for the interaction between SM Higgs and nucleon.  The spin-independent DM-nucleon scattering cross section for $m_{H_1} \gg m_h$ is evaluated as~\cite{Cline:2013gha}:
\begin{equation}
\sigma_{\chi_R-n} = \frac{\sin^2\theta_h \cos^2\theta_h}{ \pi} \frac{\mu_{nX}^2}{M_X^2} \frac{\mu^2 m_n^2 f_N^2}{v^2m_h^4 } 
\simeq 5.3 \times 10^{-43} \left( \frac{\mu \sin \theta_h \cos \theta_h}{M_{X}} \right)^2 \ {\rm [cm^2]},
\end{equation} 
where $\mu_{nX} = m_n M_X/(m_n + M_X)$ is the reduced mass, with $f_N \simeq 0.287$  for the neutron-DM scattering~\cite{Belanger:2013oya} (proton-DM scattering is almost same).  The most stringent constraint comes from XENON1T data~\cite{Aprile:2017iyp,Aprile:2018dbl} which gives $90 \%$ confidence level upper limit on $\sigma_{SI}$,  consistent with the looser bound from LUX~\cite{Akerib:2016vxi} or PandaX-II~\cite{Cui:2017nnn}.  This bound fixes the ratio of $\frac{\mu \sin \theta_h \cos \theta_h}{M_X}$ and  is recasted  into the allowed region of $(M_X, \mu)$  as shown in the right plot of Figure~\ref{fig:relic&direct}.  Based on that we can investigate the DM production via $g g  \to H_1 \to X X$ for two limits where $\sin \theta_h$ is either small or sizable.    Considering  a benchmark point of  $M_X = 200$ GeV,  the bound  $\sigma_{\chi_R-n} \lesssim 1.78 \times 10^{-46} {\rm cm}^2$ leads to   $\mu \sin \theta_h \cos \theta_h \lesssim 3.6$ GeV.  For  $\sin \theta_h \sim 0.04 $ and $\mu < 90$ GeV,  we have  $BR(H_1\to XX) < 0.3$  as indicated by Figure~\ref{fig:BRH1}, but a very small  $\sigma_{H_1}^{ggF} $ for  $m_{H_1} = 500{ \rm GeV} $  due to almost vanishing mixing.  While for a sizable  $\sin \theta_h =0.1$,  we find that at the  $\sqrt{s}=13$ TeV LHC  $\sigma_{H_1}^{ggF} (m_{H_1} = 500 {\rm GeV} )\sim \mathcal{O}(10) $ fb, but in such case   $BR( H_1 \to XX) < 0.05$ is too small since we  require $\mu < 36$ GeV.  Thus we can conclude that the DM production rate via $H_1$ exchange is negligible in this model.

\begin{figure}[t]
\centering
\includegraphics[width=0.455\textwidth]{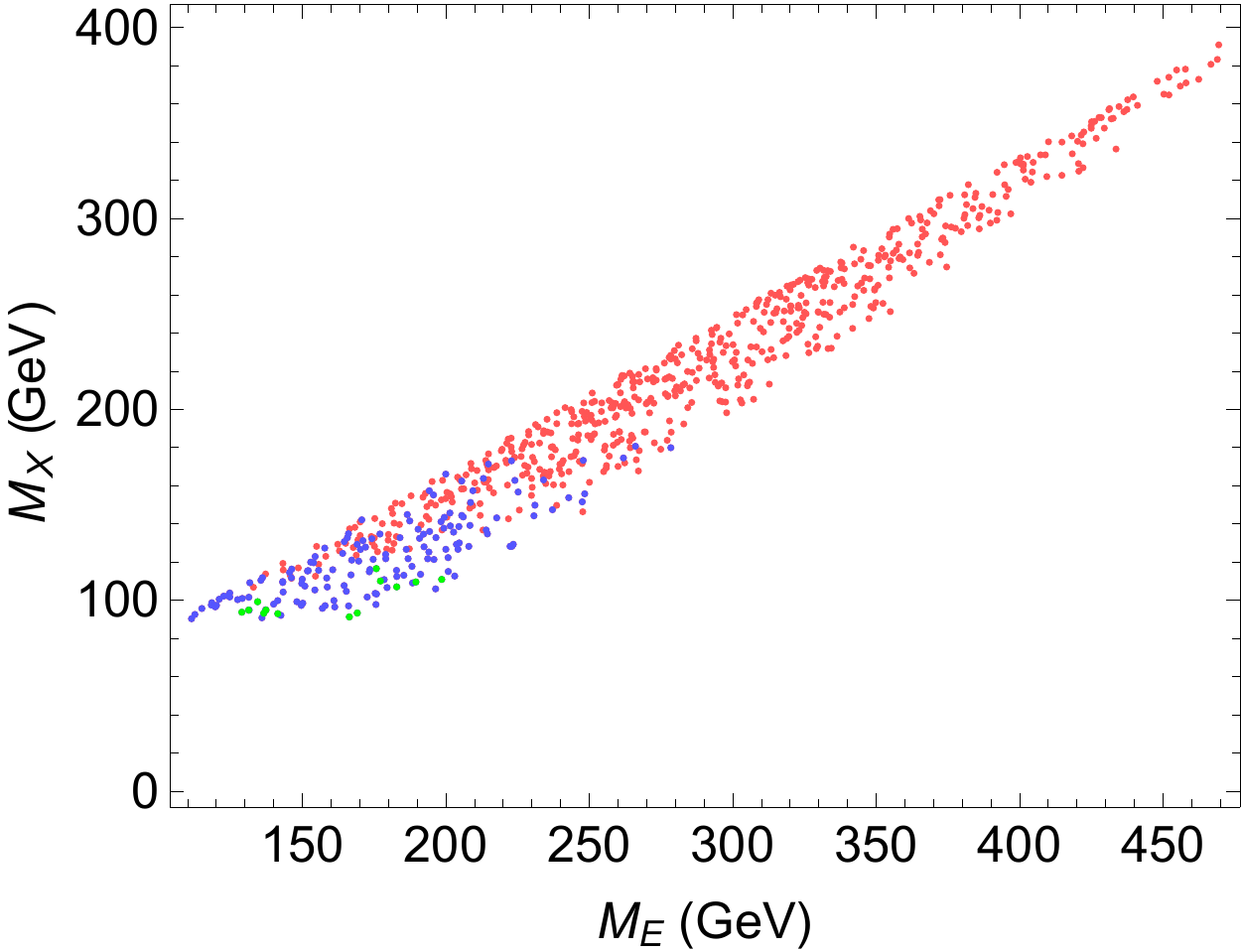}
\includegraphics[width=0.49\textwidth]{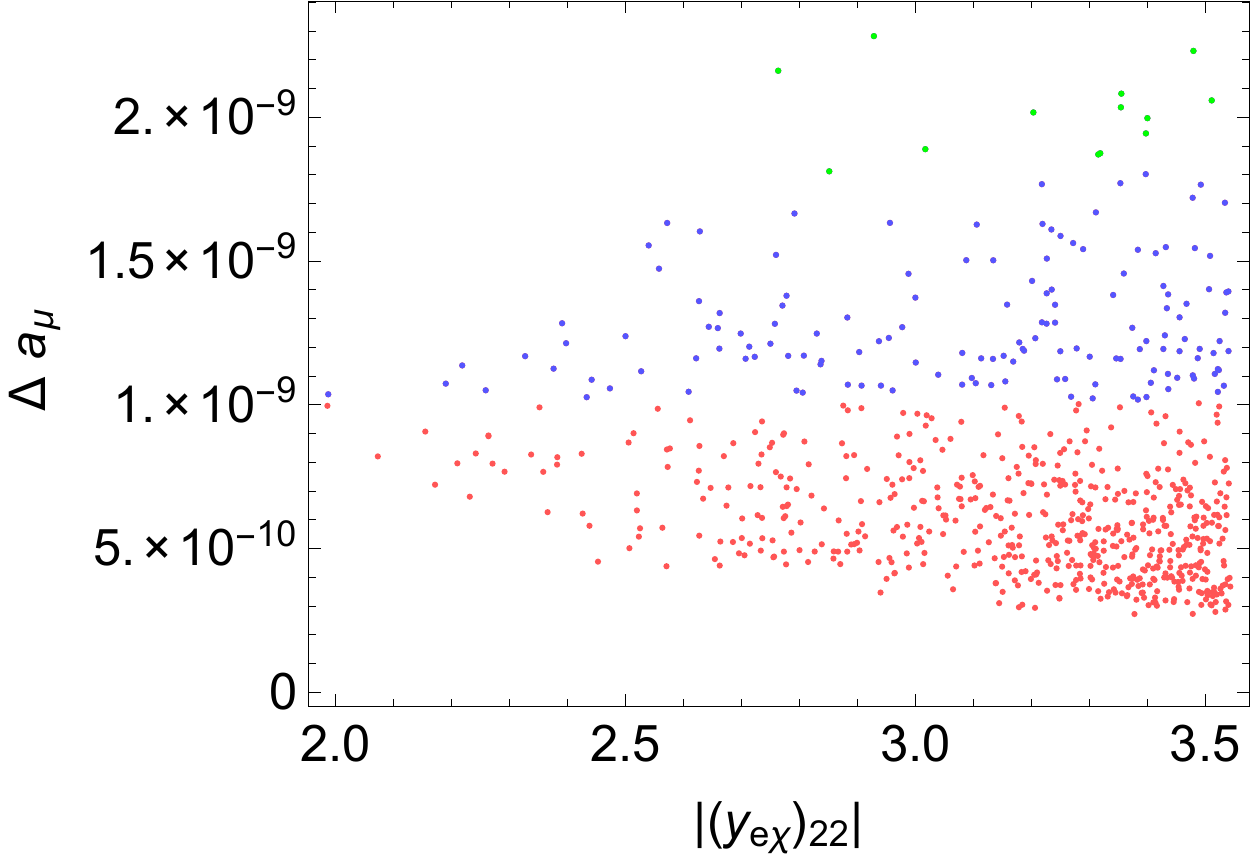}
\caption{
Left plot: The allowed region for $(M_E, M_X)$  in order to obtain  $\Delta a_\mu=(26.1\pm8.0x)\times 10^{-10}$,  $\Omega h^2=0.12 \pm0.003$ and satisfy various LFVs, where $x$=1,2,3, which is the confidence level, corresponding to green, blue, and red, respectively. The plot suggests that $M_E\lesssim 450$ GeV and $M_X\lesssim 400$ GeV. Right plot:  The correlation between  $(y_{u\chi})_{22}$ and $\Delta a_\mu $  shows that the red points actually  stand for smaller $\Delta a_\mu$ deviation.
}
\label{fig:relic&flavor}
\end{figure}  
 
 \subsection{Numerical analysis} 
Now we can combine all the bounds and  carry out  a numerical scan to  find out the parameter space which can explain  DM relic density and muon $g-2$. In this  analysis, we  show  the correlation between  $M_E$ (the lightest VLL mass) and $M_X$  by recasting the bounds from  the observed relic density and  various leptonic flavor constraints. We take the upper limit  of Yukawa couplings as $\sqrt{4\pi}$, and the regions of $M_X$, $m_{\chi_I}$ $M_{U,D}$, and $M_E$ are scanned  in the regions of $(90, 500)$ GeV, $(1.2 M_X,550)$ GeV, $(1000,2000)$ GeV, and $(1.2 M_X,1000)$ GeV respectively. Here the lower bound of $m_{\chi_I}$ is set to forbid the co-annihilation modes between $X$ and $\chi_I$ for simplicity, and   the lower limit of vector-like lepton $M_E > 108$ GeV   complies with  the  LEP experiment, although the relevant LHC limit can  be more stringent. The left plot in Figure~\ref{fig:relic&flavor} represents the allowed regions for $(M_E, M_X)$, which are consistent with  precise observations of  $\Delta a_\mu=(26.1\pm8.0)\times 10^{-10}$ and $\Omega h^2=0.12\pm0.003$, as well as satisfy  LFV and Z decay bounds. We adopt different colours in the plot to emphasize the experimental constraint from the muon g-2  at the confidence level of $68\%$ (green), $95\%$ (blue), $99.7\%$ (red).  The LFV bounds specifically  lead to the consequence  that  the typical value of $(y_{e\chi})_{22}$ should be 2$\sim$3 as verified by the right plot in Figure ~\ref{fig:relic&flavor}  and the other Yukawa couplings can be less than 1. While the masses of exotic heavy quarks $M_{U, D}$  are not  so much restricted in this simplified model.  In particular  Figure~\ref{fig:relic&flavor} indicates   that  the upper bounds for  DM and vector-like leptons masses are required to be  $M_E \lesssim 450 $ GeV and $M_X \lesssim 400$ GeV  respectively,  while the mass splitting between these two particles tends  to be small,  roughly  in a  scale of  $\sim 50$ GeV.

\section{LHC   phenomenology}
 
In this sector, we proceed to  provide an analysis  for the LHC constraint  by scanning over  the mass  region allowed by  the bounds of  flavour and relic density.  Due to the $Z_2$ parity presented in this model, exotic fermions $U$, $D$ and $E$ can  be pair produced.  In order to interpret  the LHC measurement in this  hidden $U(1)$ symmetry model, we only consider the  Drell-Yan production of  vector-like lepton (VLL) pairs, with the subsequent decaying of  $E_a \to X/\chi_{I} +  l_i $. For estimation,  the mass difference of  $( m_{\chi_I} - M_X )$ will be ignored.  The final state of   $\tau$ lepton pairs  plus  $\met$ was  recently  adopted by the CMS collaboration to extract  the  upper limit of cross section for $\tau$ slepton pair productions~\cite{Sirunyan:2018vig}.  By recasting the CMS analysis into our DM scenario, we obtain  a  loose bound for  $(M_E, M_X)$ under the assumption of universal $E_a$-$\chi$-lepton couplings, i.e. $y_{E\chi}^{e,a} = y_{E \chi }^{\mu,a} = y_{E \chi}^{\tau,a}$.

In order to simulate the $2 \tau_h + \met$ signal in this model, we employ { \sc{MG5\_aMC@NLO} }~\cite{Alwall:2014hca} to generate  events for the production of $p p \to Z, \gamma$ $ \to E^+ E^-$ at the leading order precision,  with the VLL decay into $ \chi + \{e, \mu, \tau\}$ handled by {\sc{MadSpin}}. The events are passed through {\sc{Pythia 8}}~\cite{Sjostrand:2007gs} for parton  shower and hadronization, where the $tau$ lepton decays in both leptonic and hadronic modes are  sophisticatedly processed. Event reconstruction is finally  performed by {\sc{Madanalysis 5}} package~\cite{Conte:2012fm},  so that  the jets are clustered  using  the  anti-$k_T$ algorithm implemented in {\sc{FastJet}},  with  $p_T >  20 $ GeV and a distance parameter of $R = 0.4$.  The CMS discriminant  for  $\tau_h$  reconstruction results in an  efficiency  $ \sim 60 \%$, which is also counted in our simulation.  The  event analysis is  conducted first by  a baseline selection,  demanding  two hadronic taus in opposite signs, with  a veto for electrons or muons in the final state. Subsequent kinematic cuts are  applied afterwards, including the  $M_{T2}$ variable,  sum of transverse mass $\Sigma M_T(\tau_i)$,  missing  energy $\met$ and  $\Delta \phi (\tau_1, \tau_2)$,  in order to optimize  the signal and  suppress the SM background.  The $M_{T2}$ variable  is  a  generalization of  transverse mass  into the case with two invisible particles ~\cite{Barr:2003rg, Cheng:2008hk}. In this analysis we  use the CMS interpretation  by setting the trial mass $\mu_X $ of two missing particles to be zero for a direct comparison purpose.  We  calculate the $M_{T2}$ as the minimum of all possible  maximum of $\left(M_T(\tau_1), M_T(\tau_2)\right)$,  with the partition of missing momentum in two DMs added up to be $\met$ measured  in the event:
\begin{equation}
M_{T2} = \min_{\not{p}_T^{X_1} + \not{p}_T^{X_2}= \not{p}_T }
  \left[ \max  \left( M_T(  p_T^{\tau_1}, \, \mpt^{X_1} ; \,\mu_X),\,
    M_T (  p_T^{\tau_2}, \,  \mpt^{X_2} ; \,\mu_X)  \right) \right],
\label{eq:MT2}
\end{equation}
where the transverse mass in the case of massless particles is defined as:
 \begin{equation}
 M_T (  p_T^{\tau_i}, \, \mpt^{X_i} ) =  \sqrt {2 ( E_T^{\tau_i} \met^{X_i} - p_T^{\tau_i} \cdot \not{p}_T^{X_i}) };  \, \mbox{with} ~~  i = 1, 2.
 \end{equation}
 %
 
\begin{footnotesize}
\begin{table}[t!]
\begin{center}
\begin{tabular}{|c||c|c|c|c|c|c|c|c|c||c|c|}
\hline
\multirow{2}{*}{$ \begin{array}{c} 13 ~\mbox{TeV} , 35.9~ \mbox{fb}^{-1}   \\  M_X~ \mbox{(GeV)}  \end{array} $ }&   \multicolumn{3}{|c|}{$ M_E = 150~\mbox{(GeV)} $ } & \multicolumn{3}{|c|}{$  M_E = 200~\mbox{(GeV)}  $} &  \multicolumn{3}{|c||}{$ M_E = 250~\mbox{(GeV)} $ } & \multirow{2}{*}{\footnotesize{SM BG}} &\multirow{2}{*}{\footnotesize{Observed}}
\\
 \cline{2-10}  & $~~60 ~~$& $~~80~~$ & $100$&  $~100~$  &  $~120~$  & $140$ & $ ~120~ $& $~140~ $& $160$  & &\\
\hline
$ ~40  <M_{T2} < 90 ~ \mbox{GeV}~$& $56.8$ & $62.9$ & $70.3$ & $15.8 $  & $19.1$  & $23.4$ &$5.12$ & $5.69$ & $7.08 $ & - & -
 \\
\hline
$\begin{array}{c}\Sigma M_T > 350 ~ \mbox{GeV} \\  E_T^{\mbox{miss}} > 50 ~\mbox{GeV} \end{array} $&  $5.06 $ & $3.72$ & $1.62$ & $2.35$ & $1.96$ & $1.18$ & $1.26$ & $1.23 $& $1.05 $ &- & -
\\
\hline
$\Delta \phi (l_1, l_2) > 1.5 $& $4.81$ &$ 3.40 $ & $1.43$ &  $2.27$ & $1.86$  & $1.12$ & $1.20$ & $1.19$ & $0.99$&$ 4.35^{+1.75}_{-1.53}$ & $5$
\\
\hline 
\end{tabular}
\end{center}
\caption{Number of events  after each step of selection criterion for   one generation of  VLL  with benchmark points  
$M_E = 150, 200, 250$ GeV,  for an integrated luminosity of $\int {Ldt = 35.9} ~\mbox{fb}^{-1}$  at a $\sqrt s = 13$ TeV LHC. }
\label{cuttable}
\end{table}
\end{footnotesize}
 
\begin{figure}[t!]
\centering
\includegraphics[width=0.49\textwidth]{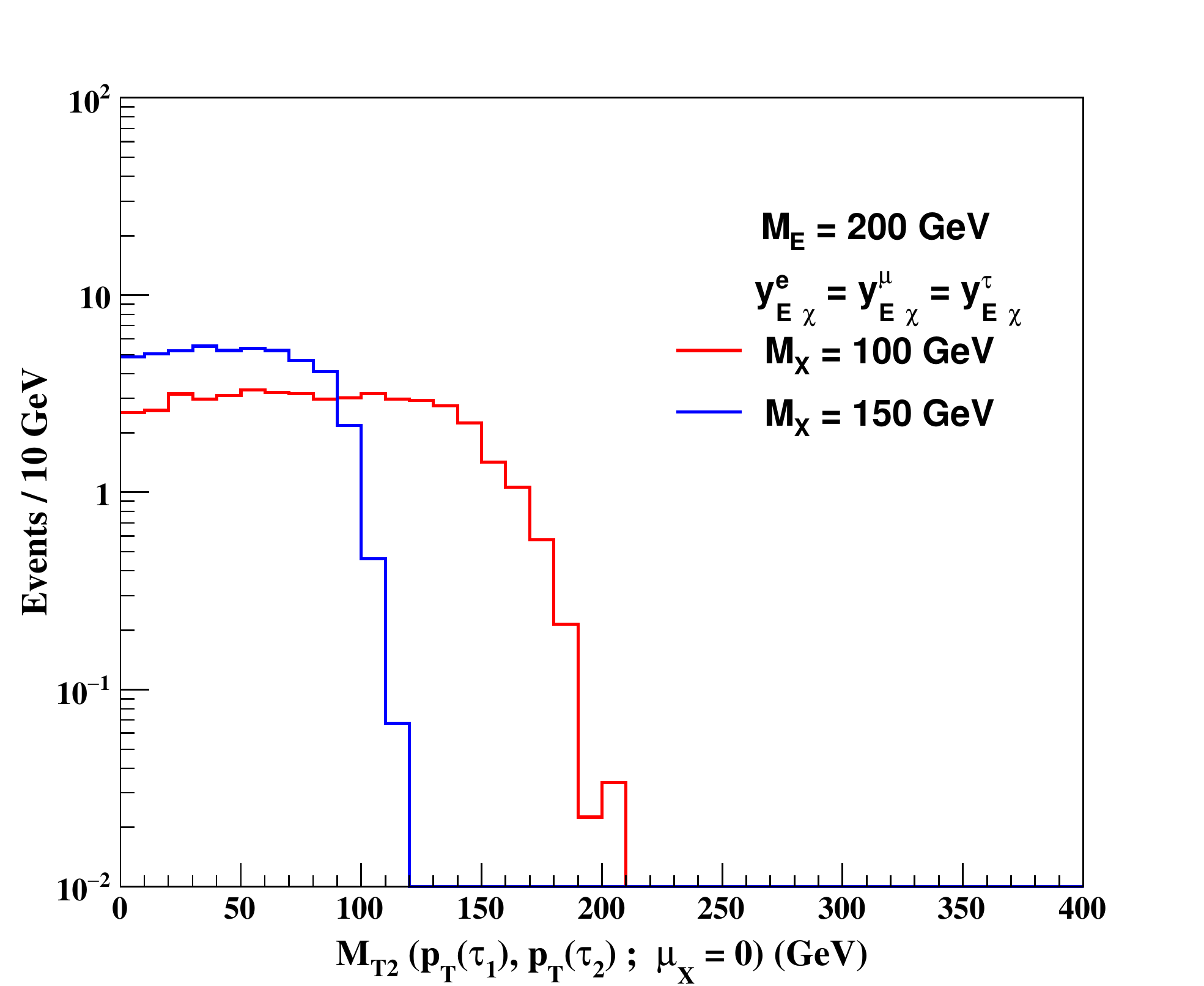}
\includegraphics[width=0.48\textwidth]{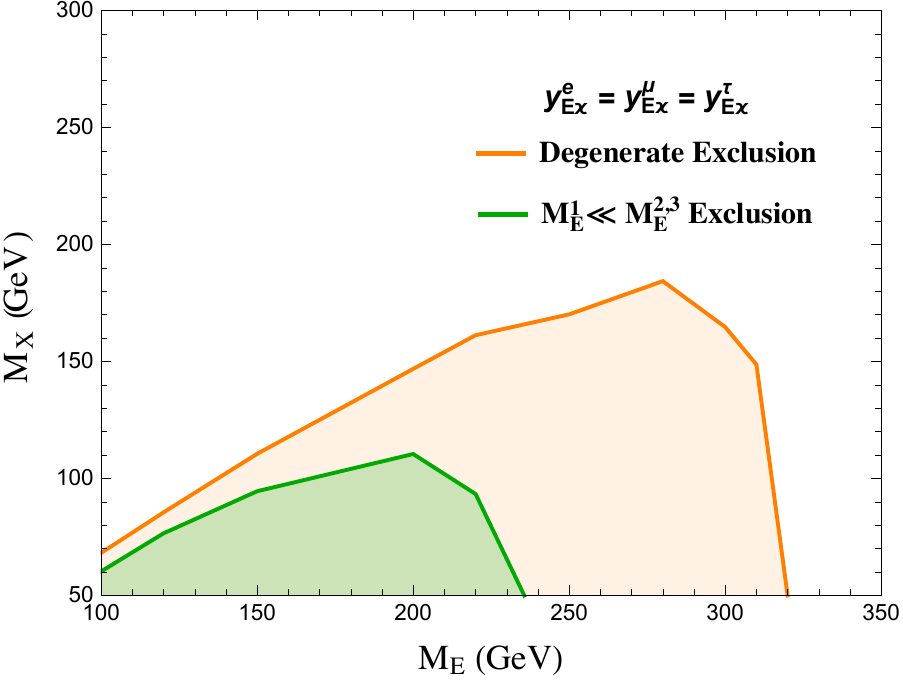}
\caption{
Left plot: The $M_{T2} $ distribution for $M_E = 200$ GeV after the baseline event selection (2  reconstructed hadronic taus  with opposite electric charge) at the $\sqrt s = 13$ TeV  LHC with a luminosity of $35.9 ~\mbox{fb}^{-1}$. Right plot:  The exclusion region  by   $2 \tau_h + \met $  measurement for universal  $E_a$-$\chi$-lepton couplings.  The green region corresponds to the  exclusion on $(M_E, M_X)$  with one  generation of VLL  mediating in  $p p \to  X X ~\tau^+ \tau^-$;  while the orange region is for  the scenario of  three generations  with degenerate  masses of $M_{E} ^1= M_{E}^{2} = M_{E}^3$.  
}
\label{fig:2tau} 
\end{figure}

Following the CMS analysis,  we  employ the event selection criteria in the search region 2 (SR2) for the $\tau_h \tau_h$ final states,  and  ignore the  selection in the other two isolated regions  of SR1 and SR3 due to their insensitivity and a larger  number of  expected SM background  than the LHC observed data. Therefore events should satisfy these requirements: (1)  $ 40 ~\mbox{GeV} <M_{T2} < 90 ~ \mbox{GeV}$, (2)   $\Sigma M_T > 350 ~ \mbox{GeV}  $, (3)   $\met  > 50 ~\mbox{GeV} $,  and (4)  $\Delta \phi (l_1, l_2) > 1.5 $.  The number of events after each cut  is reported  in the table~\ref{cuttable}, for the case  that only  the lightest  VLL is effective. The assumption of universal coupling results in equal branching ratios of $Br(E^- \to e^- + \chi) = Br(E^- \to \mu^- + \chi) = Br(E^- \to \tau^- + \chi) = 1/3$, which can be  consistent with the flavour constraint.  The cut table indicates  that for a fixed VLL mass $M_E$, the event number after those $\Sigma M_T $, $\met$ and $\Delta \phi (l_1,  l_2)$ cuts  decreases  for an increasing $M_X$. While the event number after the $M_{T2}$ cut will instead  be enhanced in that situation. This is a reflection of the $M_{T2}$ quality as a function of the trial mass for missing particle. If the trial mass $\mu_X$ equals the true mass  $ M_X$, the end point of $M_{T2}$  gives the exact mass of the parent particle $M_E$.  However  for the  large  deviation  $M_X \gg  \mu_X = 0 $,  the end point $M_{T2}$ will drop  below  $M_E$  because there is  less measured missing energy.  In the left plot of Figure~\ref{fig:2tau}, we present the $M_{T2}$ distribution after the basic cut for $M_E = 200 $ GeV,  $M_X = 100, 150$ GeV.  For a larger DM mass,  the event distribution  shifts into the lower mass region due to a false trial mass,  leading to  an increase  for the $M_{T2}$ cut acceptance.

The  CMS   collaboration provides a simulation for the SM background,  which is $4.35^{ + 1.75}_{ - 1.53}$ in the SR2 signal region, and the  observed  event numbrer is $5$  at the 13 TeV LHC.  This can be translated into a $68\%$ C.L. exclusion limit  for  $(M_E, M_X)$  presented in the right plot of Figure~\ref{fig:2tau}.  As we can see,  with only  one generation of VLL, the LHC constraint  is not stringent,  excluding  a small mass region with $M_X \lesssim 110$ GeV for  $M_E \sim 200$ GeV.  While for the three-generation scenario, the exclusion becomes much more relevant. The  upper exclusion limit for $M_X$  reaches $185$ GeV, which possibly overlaps with the mass region permitted by the  relic density and  flavor bounds  displayed in Figure~\ref{fig:relic&flavor}.  Note that all three generations of VLLs  contribute to this specific LHC signal with a mass hierarchy of $M_{E1} < M_{E2} < M_{E3}$. However since the Yukawa coupling $(y_{e\chi})_{22}$ is preferred to be larger than other ones,  the assumption of universal couplings is less observed for the second generation.  Thus  the realistic LHC exclusion region for  $(M_E, M_X)$  would  most likely lie below the upper  boundary of the orange band.

\section{Summary and Conclusions}
We have proposed  an inverse seesaw scenario in a framework of hidden $U(1)_H$ gauge symmetry  where  extra scalars  and vector-like  neutrinos are introduced to assist  the mass generation of neutrino,  while  vector-like quarks and leptons are required  in order to  cancel the $U(1)_H$ gauge anomaly.  For the neutrino sector,  we apply the  Casas-Ibarra parametrisation  to fit  neutrino oscillation data and bound of non-unitary PMNS matrix.   This model features a bosonic dark matter candidate stabilised by  a $Z_2$ parity.  Specifically  the DM interaction with  exotic charged fermions  plays an important role to realize the observed relic density in a viable  limit $v_{H'} \ll v_H <  v_{\varphi'} \ll v_{\varphi}$. 
By tuning  the scalar potential,  a  minimal mixing among the SM Higgs and extra scalars is  achieved. Under this assumption, we  provide the allowed region  capable  to accommodate  the discrepancy in  muon $g-2$,  while  consistent with the relic density and  flavour bounds.  In case that a  notable  $h-\varphi'$ mixing  is  invoked,  the  Higgs portal DM-nucleon scattering fixes the upper limit for   $\mu \sin \theta_h \cos \theta_h/M_X$.  Our analysis shows that after  taking into account the  constraint from direct detection, the cross section of DM  pair production via a heavy Higgs $H_1$ is almost negligible.

Concerning the possibility to extract the DM mass bound at the LHC, we  focus on the Drell-Yan  pair production of vector-like charged lepton since  it provides clear signal  of  charged leptons plus missing transverse energy involving DM.  We recasted the  CMS analysis for   $2\tau_h + \met$ events  at the $\sqrt{s}= 13$ TeV LHC based on the $M_{T2} $ selection, which shows  that the lower regions in  ($M_E$, $M_X$)  are ruled out  depending on the mass degeneracy among vector-like charged leptons and their branching ratios into tau leptons.  However due to the current  insensitivity to $\tau_h$, most of the allowed region from  relic density and flavor physics would survive for  Yukawa couplings in correct orders. \\

\section*{Acknowledgments}
This research is supported by the Ministry of Science, ICT \& Future Planning of Korea, the Pohang City Government, and the Gyeongsangbuk-do Provincial Government (H. C. and H. O.).


\end{document}